\documentclass{cernrep}
\usepackage{texnames}
\usepackage[T1]{fontenc}
\usepackage{epstopdf}

\usepackage{xfrac}

\begin{document}
\title{One-Quadrant Switched-Mode Power Converters}

\author{R. Petrocelli}

\institute{ALBA Synchrotron, Barcelona, Spain}

\maketitle 

\begin{abstract}
This article presents the main topics related to one-quadrant power converters. The basic topologies are analysed and a simple methodology to obtain the steady-state output--input voltage ratio is set out. A short discussion of different methods to control one-quadrant power converters is presented. Some of the reported derived topologies of one-quadrant power converters are also considered. Some topics related to one-quadrant power converters such as synchronous rectification, hard and soft commutation, and interleaved converters are discussed. Finally, a brief introduction to resonant converters is given.\\\\
{\bfseries Keywords}\\
One-quadrant; converter; topology; switch-mode; magnet; energy.
\end{abstract}

\section{Introduction}

Switched-mode power converters are a very efficient way to transfer energy from a source to a load. A switched-mode power converter is formed from switches, inductors, and capacitors. If these components are ideal, they do not dissipate energy, and the efficiency of the power converter is 100\%.
A further advantage of switched-mode power converters is the increase in power density that has occurred in recent years. The power density is defined as the ratio between the nominal power of the converter and its volume. This power density has been constantly increasing owing to an increasing switching frequency.

As this is a broad topic, there are several different ways to classify switched-mode power converters. The one-quadrant power converters are those with the capability to provide an output in only one quadrant of a plot of output voltage versus output current. These converters have the capability to transfer energy in only one direction. Most one-quadrant converters can also be classified as DC--DC converters. The main application of these converters is in voltage regulation. In this application, a non-regulated voltage, which is usually provided by a diode rectifier stage, is converted to a regulated output voltage which is not affected by the grid or variations in the load.

This article gives an introduction to one-quadrant power converters. It presents the basic, or direct, DC--DC converter topologies, showing typical waveforms and the basic design equations. The basic concepts of regulation and control of DC--DC converters are also discussed.
Section \ref{sec:derived} presents some of the most typical derived converter topologies reported in the literature.
Some topics related to one-quadrant power converters, such as synchronous rectification, interleaving, and soft commutation, are presented in Section \ref{sec:topics}. Finally, resonant converters are briefly  mentioned.

\subsection{One-quadrant switched-mode power converters in particle accelerators}
The main application of one-quadrant power converters is in storage ring particle accelerators. These accelerators are the key components of synchrotron radiation light source facilities. A storage ring is a particle accelerator where the particles circulate around a ring for several hours while their energy is maintained constant.
As the energy of the particles is constant, the magnetic fields used for bending
 and focusing the particle beam are also constant. These magnetic fields are generated by currents provided by high-precision, stable power supplies. The precision and stability of these converters are of the order of several parts per million.

\section{Basic one-quadrant power converter topologies}

This section provides a short introduction to the basic topologies of one-quadrant power converters. These basic topologies are also known as direct DC--DC converters. One characteristic of these topologies is that they are built using only one commutation cell.
A brief description, a methodology for the analysis of these topologies, and key design equations are presented.
First, the DC chopper converter is discussed as an introduction and to show how the average output voltage of a converter can be modified using `on--off' switches. The basic topologies considered are the buck, boost, and buck--boost converters.

\subsection{DC chopper}
\begin{figure}
\centering
\includegraphics{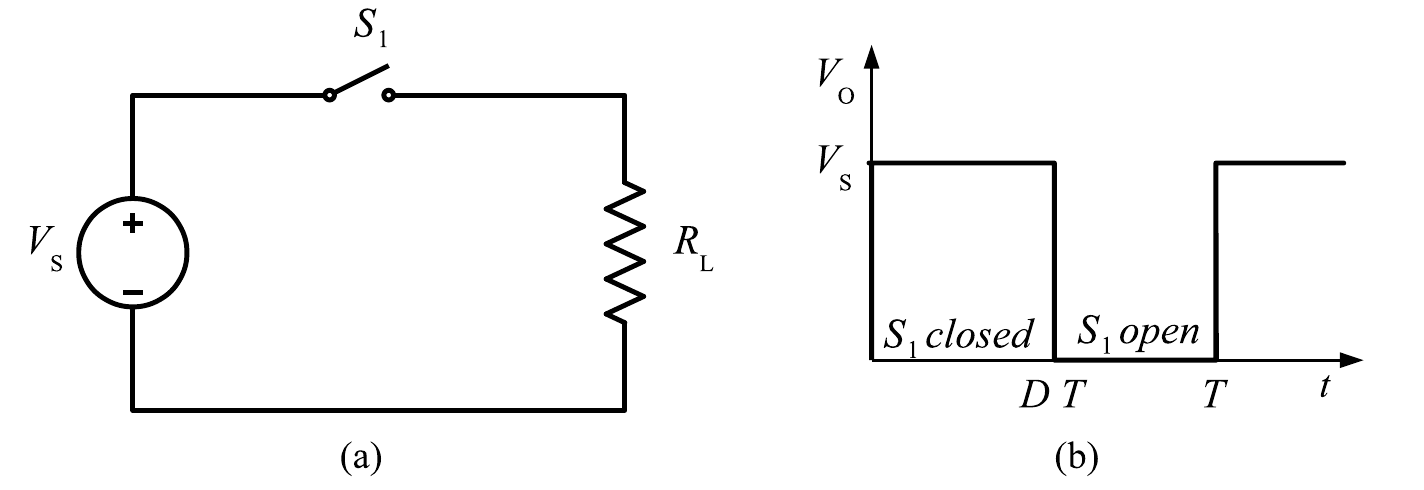}
\caption{Simple DC chopper: (a) circuit diagram; (b) output voltage waveform}
\label{fig:chopper01}
\end{figure}

The simplest one-quadrant converter is the DC chopper, shown in \Fref{fig:chopper01}. This consists of a DC input voltage source, a controllable switch, and a load resistor. When the switch is closed, the input voltage is applied to the resistor. This is defined as the `on' state of the converter. When the switch is open, the current through the resistor is zero and the voltage is zero. This is called the `off' state of the converter.

The switch can be operated with a duty ratio $D$, defined as the ratio of the on time of the switch to the sum of the on and off times. For constant-frequency operation,
\begin{equation}
	D  = \frac{t_\mathrm{on}}{t_\mathrm{on} + t_\mathrm{off}} \frac{t_\mathrm{on}}{T},
\label{eq:chopper01}
\end{equation}
where $T=1/f$ is the period of the switching frequency $f$. The average value of the output voltage is
\begin{equation}
V_\mathrm{O} = D V_\mathrm{S}.
\label{eq:chopper02}
\end{equation}
This average output voltage can be regulated by adjusting the duty cycle $D$.

\begin{figure}
	\centering
	\includegraphics{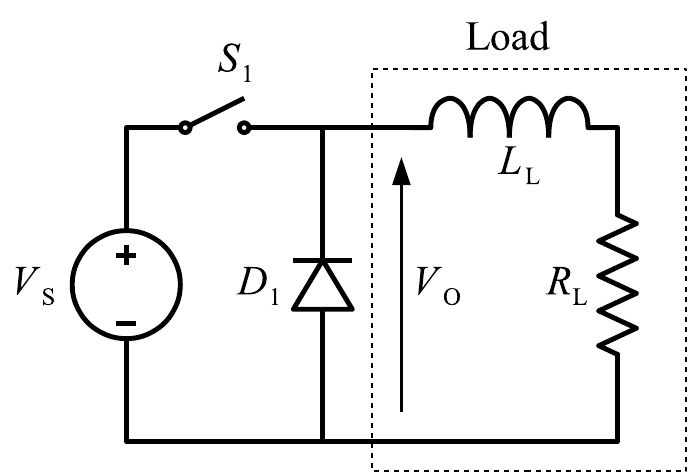}
	\caption{DC chopper with inductive load}
	\label{fig:chopper02}
\end{figure}

The circuit shown in \Fref{fig:chopper01} is not a `practical' circuit. The main reason for this is that it is not possible to build pure resistive elements. In practice, the connection of devices and circuit elements creates stray inductances that have to be taken into account. A circuit including the stray inductance of the load is shown in \Fref{fig:chopper02}.

\begin{figure}
 	\centering
 	\includegraphics{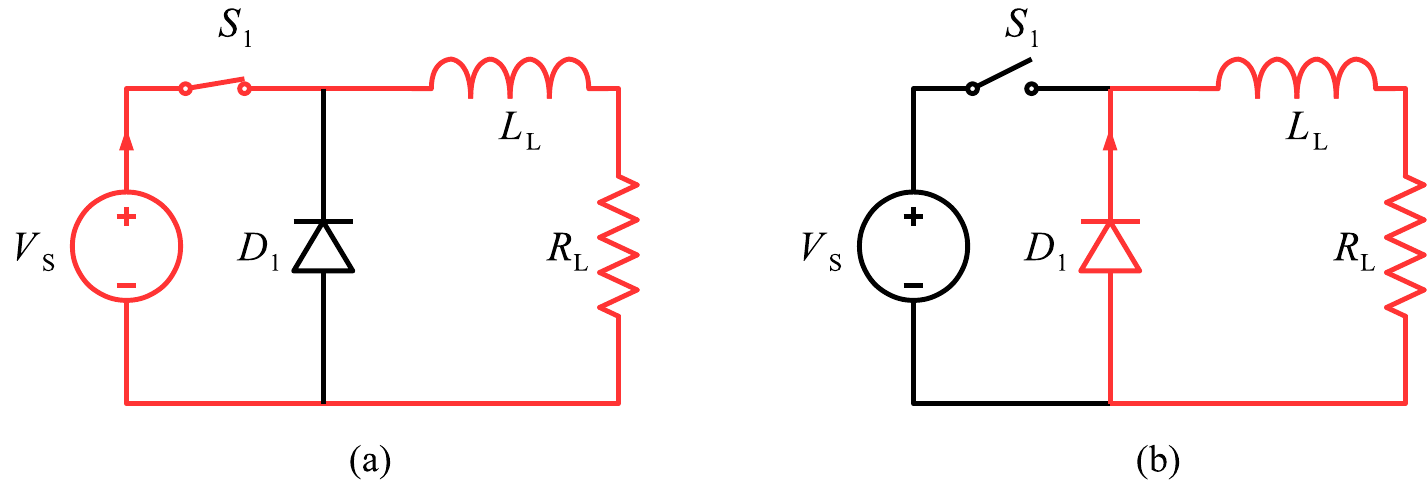}
 	\caption{DC chopper: (a) current path in on state; (b) current path in off state}
 	\label{fig:chopper03}
\end{figure}

The inclusion of the inductance of the load leads to a need to provide a path for the current in the inductor. This is to comply with the `rule' of power electronics that a current source that is like an inductor must not be open-circuited. Two circuits are defined according to the state of the switch; these circuits are shown in \Fref{fig:chopper03}. The current paths are shown in red.

\begin{figure}
	\centering
	\includegraphics{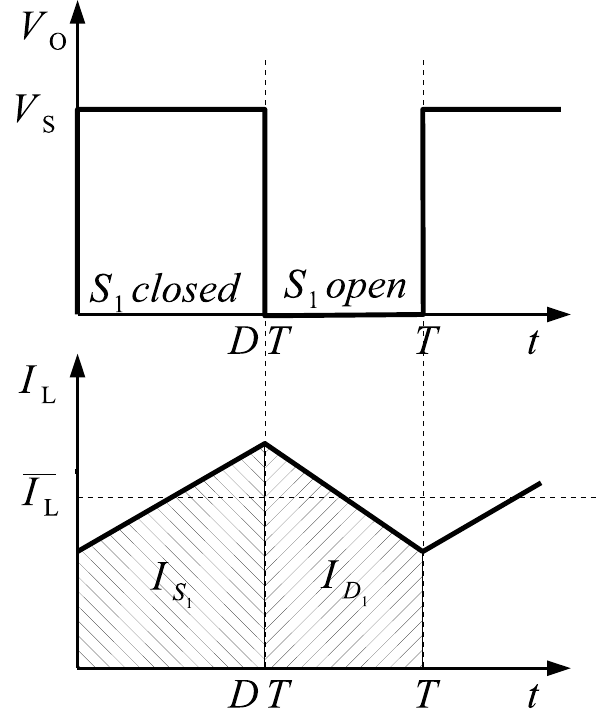}
	\caption{DC chopper: output voltage and load current}
	\label{fig:chopper04}
\end{figure}

The load voltage and current are plotted in \Fref{fig:chopper04} under the assumptions that the load current never reaches zero and the time constant of the load $\tau = {L_\mathrm{L}}/{R_\mathrm{L}}$ is much greater than the period $T$. The average values of the output voltage and current can be adjusted by changing the duty ratio $D$. This is the principle of control of switched-mode power converters.

\subsection{Buck converter}
\begin{figure}
\centering
\includegraphics{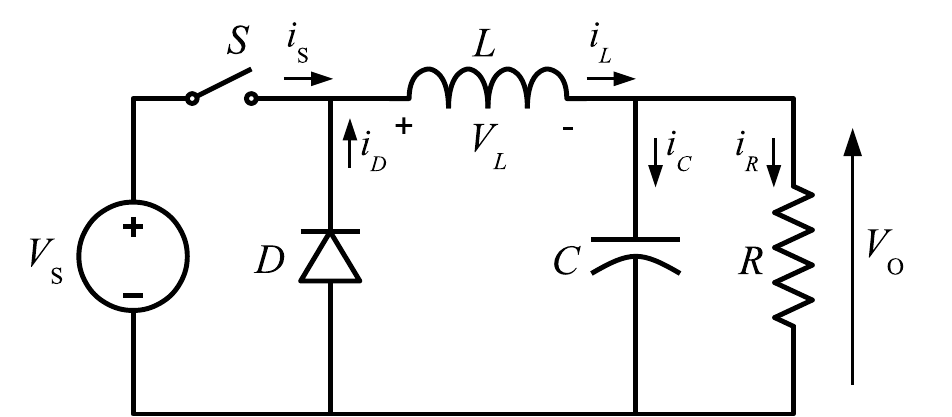}
\caption{Buck (step-down) converter}
\label{fig:buck01}
\end{figure}

The addition of a capacitor to the circuit of \Fref{fig:chopper02}
 leads to the first topology of basic, or direct, DC--DC converters. These are step-down  converters and are commonly known as buck converters. The basic circuit diagram of a buck converter is shown in \Fref{fig:buck01}. It consists of a DC input voltage source $V_\mathrm{S}$, a controlled switch $S$, a diode $D$, a filter inductor $L$, a filter capacitor $C$, and a load resistance $R$.

\begin{figure}
\centering
\includegraphics{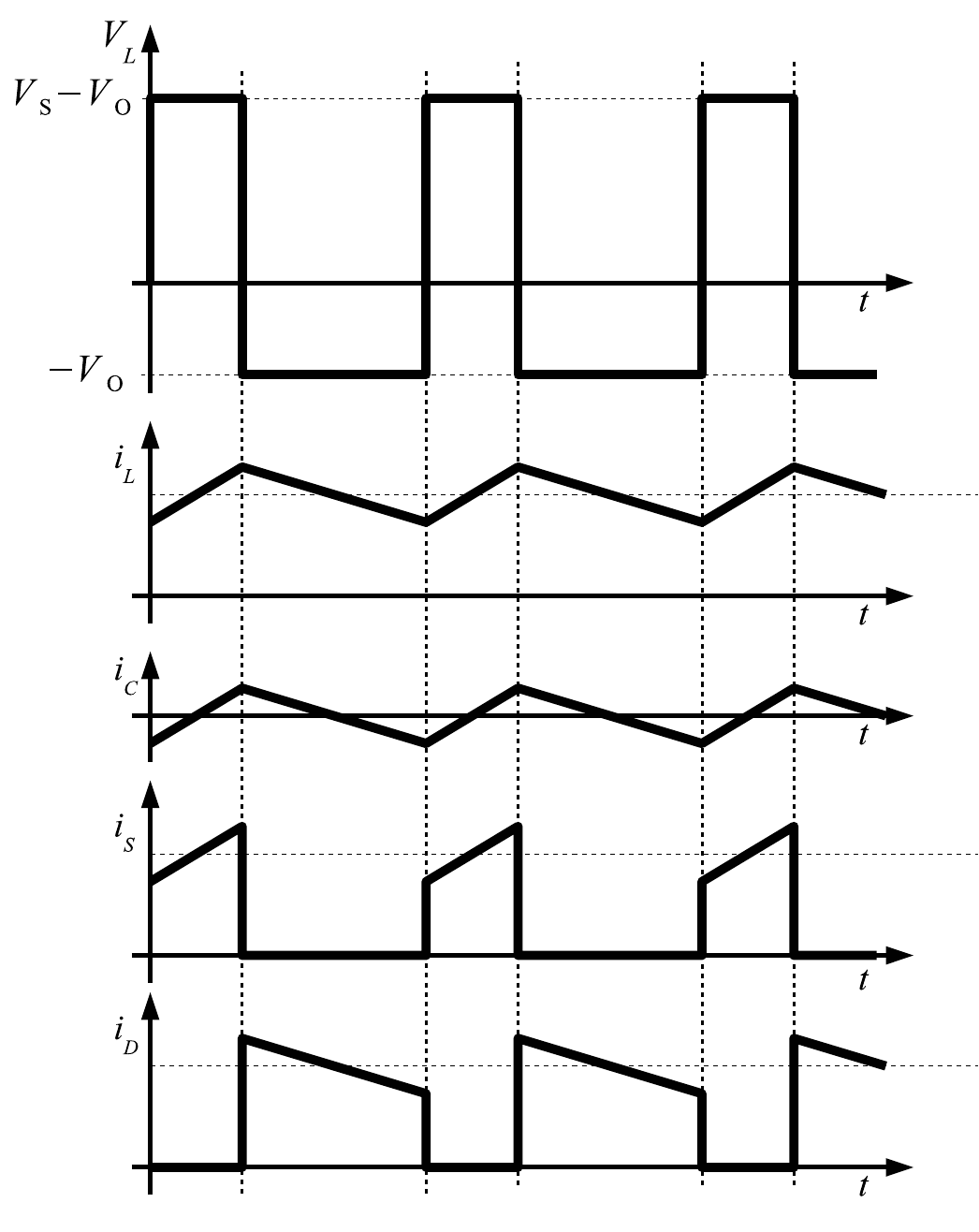}
\caption{Waveforms for buck converter}
\label{fig:buckWaveforms}
\end{figure}

DC--DC converters can operate in two distinct modes with respect to the inductor current $i_L$. Figure \ref{fig:buckWaveforms} depicts the continuous conduction mode (CCM), in which the inductor current is always greater than zero. When the average value of the input current is low (high $R$) and/or the switching frequency $f$ is low, the converter may enter the discontinuous conduction mode (DCM). In the DCM, the inductor current is zero during a portion of the switching period.

Typical waveforms of the converter in the CCM are shown in \Fref{fig:buckWaveforms}. It can be seen from the circuit that when the switch $S$ is commanded to the on state, the diode $D$ is reverse biased. When the switch $S$ is off, the diode conducts so as to support an uninterrupted current in the inductor.

The relationships among the input voltage, the output voltage, and the switch duty ratio $D$ can be derived from, for instance, the waveform of the inductor voltage $V_L$ (\Fref{fig:buckWaveforms}). According to Faraday's law, the volt-second product for the inductor over a period of steady-state operation should be zero. For the buck converter,
\begin{equation}
(V_\mathrm{S} - V_\mathrm{O}) D T = V_\mathrm{O} (1-D) T.
\label{eq:buck01}
\end{equation}
From Eq. (\ref{eq:buck01}), the steady-state DC voltage transfer function, defined as the ratio of the output voltage to the input voltage, is
  \begin{equation}
M_\mathrm{V}\equiv\frac{V_\mathrm{O}}{V_\mathrm{S}} = D.
\label{eq:buck02}
\end{equation}
It can be seen from Eq. (\ref{eq:buck02}) that the output voltage is always smaller than the input voltage.

The CCM is preferred over the DCM because it provides higher efficiency and makes better use of semiconductor switches and passive components. The DCM may be used in applications with special control requirements, since the dynamic order of the converter is reduced (the energy stored in the inductor is zero at the beginning and end of each switching period). It is uncommon to mix these two operating modes, because different control algorithms are needed. For the buck converter, the critical value of the inductance which is needed for CCM operation is
\begin{equation}
L_\mathrm{b} = \frac{(1-D)R}{2 f}.
\label{eq:buck03}
\end{equation}

The current $i_L$ in the filter inductor in the CCM consists of a DC component $I_\mathrm{O}$ with a superimposed triangular AC component. Almost all of this AC component flows through the filter capacitor as a current $i_C$. The current $i_C$ causes a small voltage ripple in the DC output voltage $V_\mathrm{O}$. To limit the peak-to-peak value of the ripple voltage below a certain value $V_\mathrm{r}$, the filter capacitance $C$ must be greater than
\begin{equation}
C_\mathrm{MIN} = \frac{(1-D)V_\mathrm{O}}{8 V_\mathrm{r} L f^2}.
\label{eq:buck04}
\end{equation}

Equations (\ref{eq:buck03}) and (\ref{eq:buck04}) are the key design equations for the buck converter. The input and output DC voltages (and, hence, the duty ratio $D$) and the range of the load resistance $R$ are usually determined by preliminary specifications. The designer needs to determine the values $L$ and $C$ of the passive components, and the switching frequency $f$. The value $L$ of the filter inductor is calculated from the CCM/DCM condition using Eq. (\ref{eq:buck03}).

The value $C$ of the filter capacitor is obtained from the voltage ripple condition (Eq. (\ref{eq:buck04})). For compactness and low conduction losses in the converter, it is desirable to use small passive components. Equations (\ref{eq:buck03}) and (\ref{eq:buck04}) show that this can be accomplished by using a high switching frequency $f$. The switching frequency is limited, however, by the type of semiconductor switches used and by switching losses. It should also be noted that values of $L$ and $C$ may be altered by the effects of parasitic components in the converter, especially the equivalent series resistance of the capacitor.

\subsubsection{Discontinuous current mode}
\begin{figure}
\centering
\includegraphics[scale=0.9]{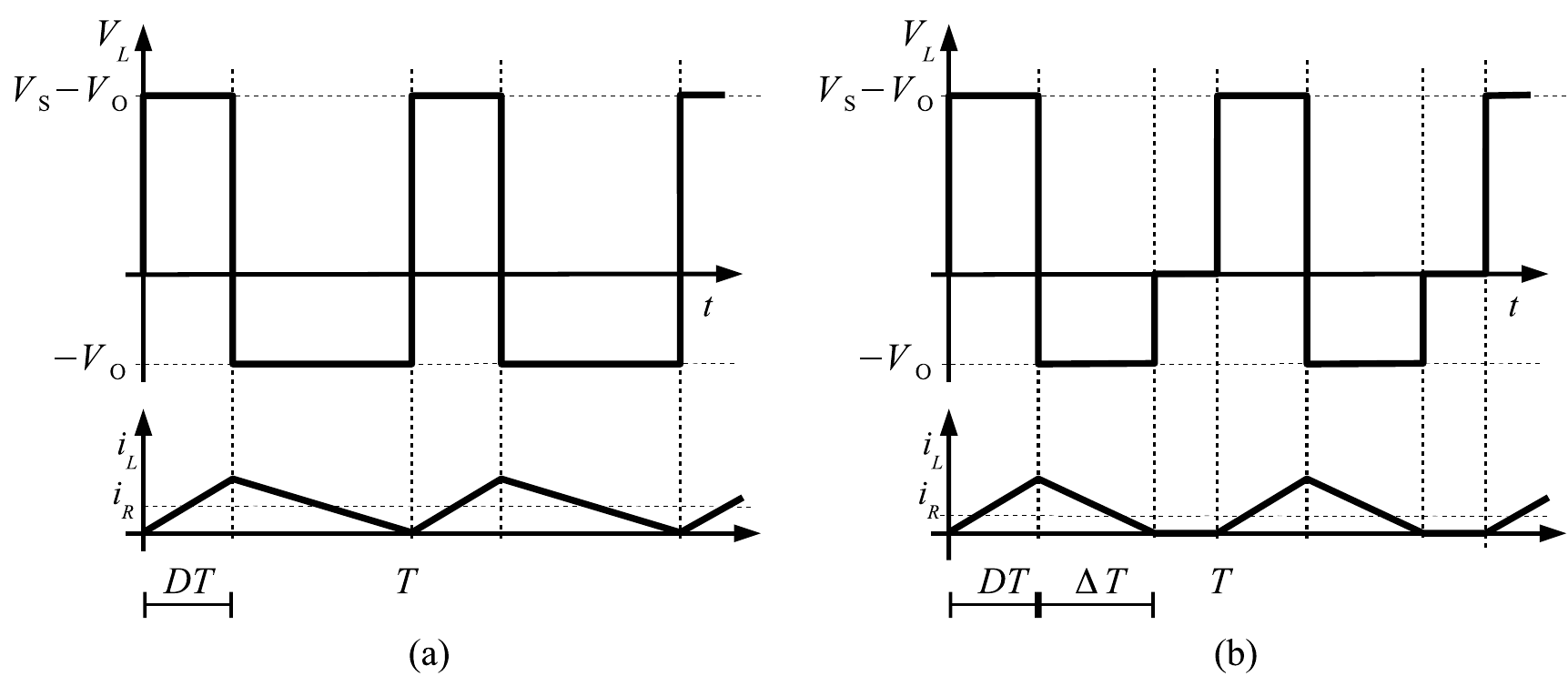}
\caption{CCM and DCM: (a) boundary condition waveforms; (b) DCM waveforms}
\label{fig:buck_DCM01}
\end{figure}

Whether the converter operates in the CCM or the DCM depends on the load on the converter. If the value of the output resistance increases beyond the value used in Eq. (\ref{eq:buck03}), the converter starts to operate in the DCM.
Figure \ref{fig:buck_DCM01}(a)
 shows the waveforms of the inductor voltage $v_L$ and inductor current $i_L$ at the edge of the CCM. The value of the output current $i_R$ is also shown in the plot. This value is the average of the inductor current over a period $T$.

\begin{figure}[h]
\centering
\includegraphics{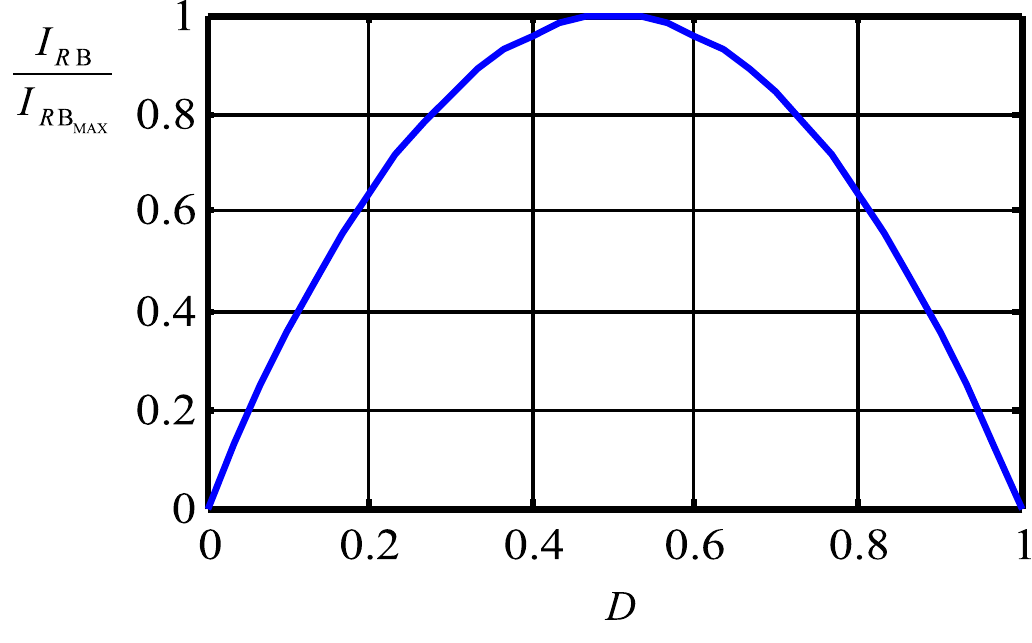}
\caption{Minimum output current for CCM versus duty cycle keeping $V_\mathrm{S}$ constant}
\label{fig:buck_DCM02}
\end{figure}

The output current $i_R$  for the condition at the boundary between the two modes is given by
\begin{equation}
i_{R\mathrm{B}} = \frac{D T}{2 L} (V_\mathrm{S} - V_\mathrm{O}) = \frac{T V_\mathrm{S}}{2 L} (D-D^2).
\label{eq:buckDCM01}
\end{equation}
The output current required in the CCM is maximum for
\begin{equation}
I_{R\mathrm{B_{MAX}}} = \frac{ T V_\mathrm{S} }{ 8 L}.
\label{eq:buck_DCM02}
\end{equation}
Figure \ref{fig:buck_DCM02} shows a plot of the minimum output current for CCM operation as a function of the duty cycle $D$.

The output--input voltage ratio under DCM conditions can be calculated using the waveforms in \Fref{fig:buck_DCM01}(b). The volt-second product over a period should be zero for a steady-state condition. Therefore
\begin{equation}
(V_\mathrm{S} - V_\mathrm{O}) D T = V_\mathrm{O}  \Delta T ,
\end{equation}
\begin{equation}
\frac{V_\mathrm{O}}{V_\mathrm{O}} = \frac{D}{D \Delta} .
\label{eq:buck_DCM03}
\end{equation}
The output current is the average of the inductor current $I_L$ over the period $T$:
\begin{equation}
I_R = \frac{I_{L_\mathrm{peak}} (D + \Delta) T }{2 T}.
\label{eq:buck_DCM04}
\end{equation}
Using Eqs. (\ref{eq:buck_DCM02}) and (\ref{eq:buck_DCM04}), Eq. (\ref{eq:buck_DCM03}) can be rewritten as
\begin{equation}
I_R = 4 I_{R\mathrm{B_{MAX}}} D \Delta.
\label{eq:buck_DCM05}
\end{equation}
From Eqs. (10) and (13),
\begin{equation}
\frac{V_\mathrm{O}}{V_\mathrm{S}} = \frac{D^2}{\left( D^2 + \frac{1}{4} \frac{I_R}{I_{R\mathrm{B_{MAX}}}} \right)}.
\label{eq:buck_DCM06}
\end{equation}

\begin{figure}
\centering
\includegraphics{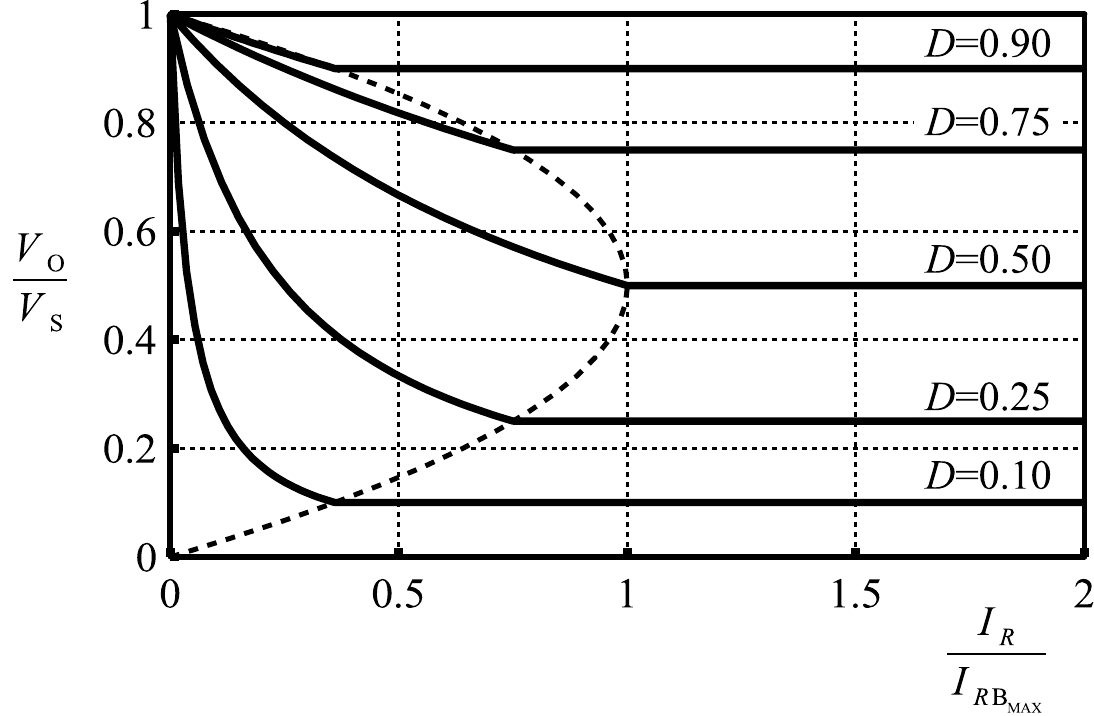}
\caption{Buck converter: output--input voltage ratio in DCM and CCM}
\label{fig:buck_DCM03}
\end{figure}

Figure \ref{fig:buck_DCM03} shows the characteristics of the buck converter for both modes of operation. The voltage ratio is plotted versus the output current for several duty cycles. The boundary between the DCM and CCM is shown by the dashed curve.

\subsection{Boost converter}

\begin{figure}[]
	\centering
	\includegraphics{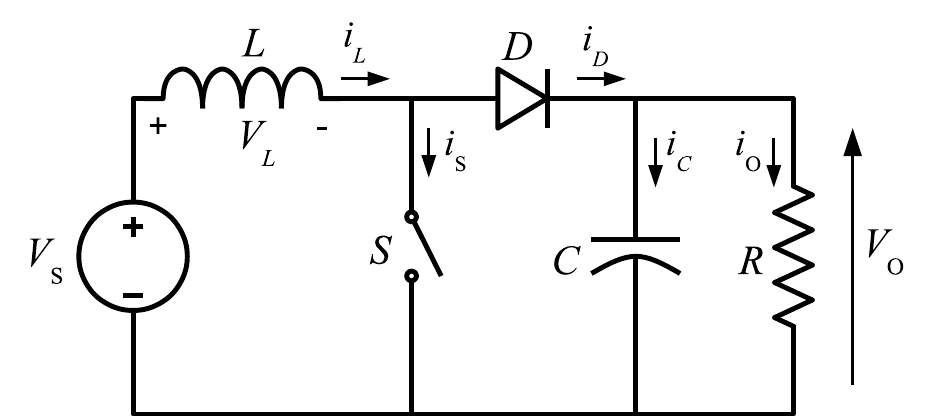}
	\caption{Boost converter: circuit diagram}
	\label{fig:boost01}
\end{figure}

\begin{figure}[h]
	\centering
	\includegraphics[scale=1]{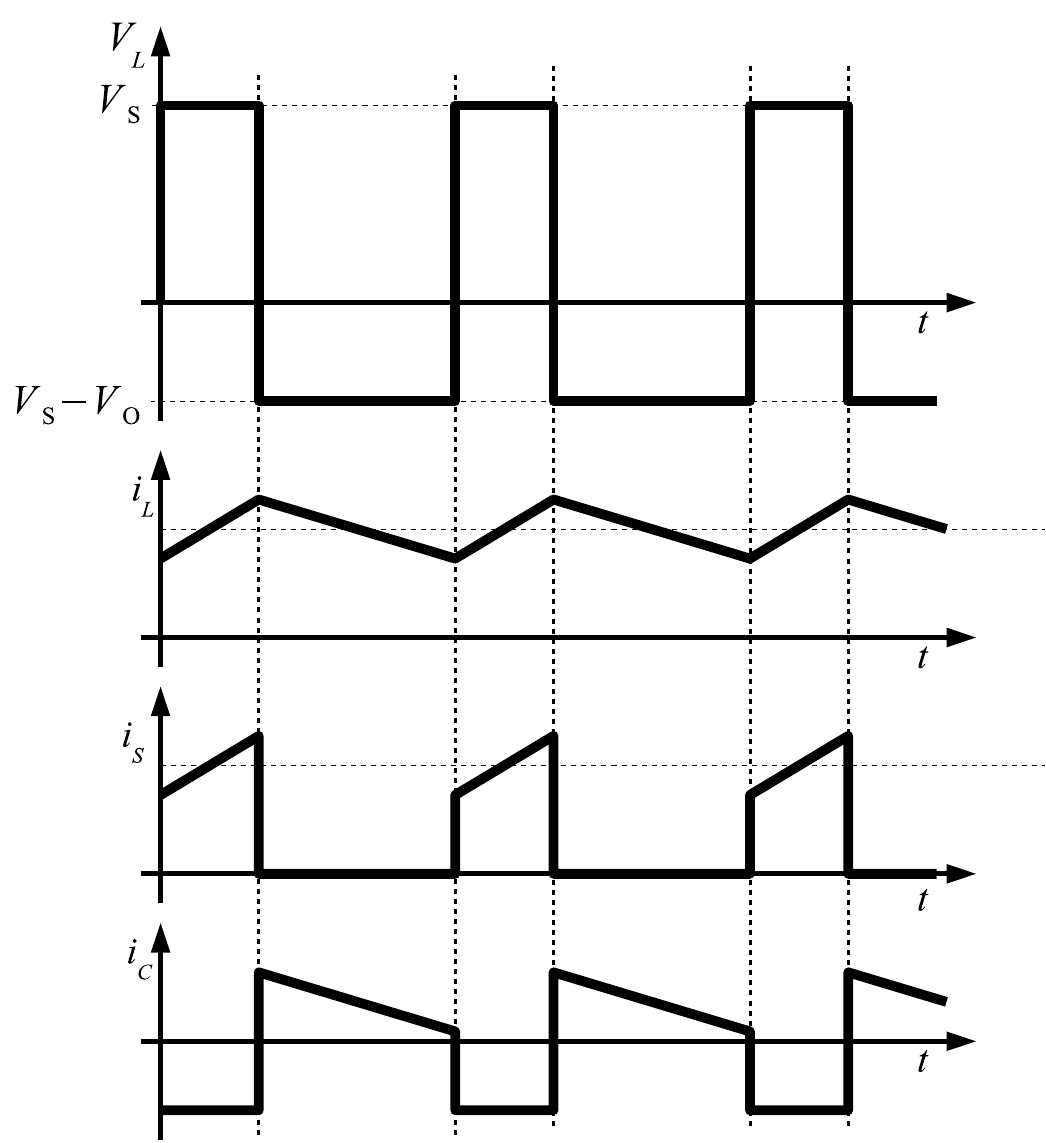}
	\caption{Waveforms for boost converter}
	\label{fig:boost02}
\end{figure}

The second topology to be analysed is the step-up, or boost, converter. It receives this name because the output voltage is always higher than the input voltage.
Figure \ref{fig:boost01} depicts a simple circuit diagram of a  boost converter. This comprises a DC input voltage source $V_\mathrm{S}$, a boost inductor $L$, a controlled switch $S$, a diode $D$, a filter capacitor $C$, and a load resistance $R$. The converter waveforms in the CCM are presented in \Fref{fig:boost02}. When the switch $S$ is in the `on' state, the current in the boost inductor increases linearly. The diode $D$ is off at this time. When the switch $S$ is turned off, the energy stored in the inductor is released through the diode into the output RC circuit.

Using Faraday's law for the boost inductor,
\begin{equation}
V_\mathrm{S} D T = (V_\mathrm{O} - V_\mathrm{S})(1-D)T,
\label{eq:boost01}
\end{equation}
from which the DC voltage transfer function turns out to be
\begin{equation}
M_\mathrm{V} \equiv \frac{V_\mathrm{O}}{V_\mathrm{S}} = \frac{1}{(1-D)}.
\label{eq:boos02}
\end{equation}
The boost converter operates in the CCM for $L>L_\mathrm{b}$, where
\begin{equation}
L_\mathrm{b} = \frac{(1-D)^2 D R}{2 f}.
\label{eq:boos03}
\end{equation}

The main waveforms of the boost converter are shown in \Fref{fig:boost02}. The current supplied to the RC output filter  is discontinuous. Thus, a larger output filter capacitor is required than in the buck converter in order to limit the output voltage ripple. The output filter capacitor must provide the DC current to the load when the diode $D$ is off. The minimum value of the filter capacitance that results in a voltage ripple $V_\mathrm{r}$ is given by
\begin{equation}
C_\mathrm{min} = \frac{D V_\mathrm{O}}{V_\mathrm{r} R f}.
\label{eq:boost04}
\end{equation}

\subsection{Buck--boost converter}
\begin{figure}
	\centering
	\includegraphics{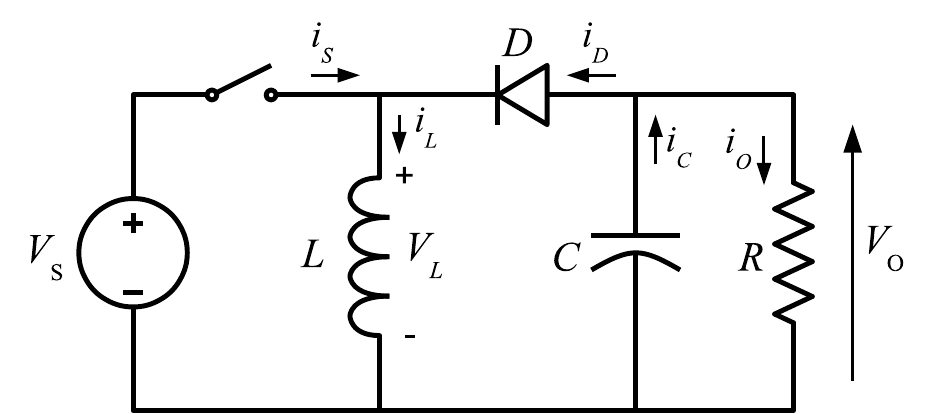}
	\caption{Buck--boost converter: circuit diagram}
	\label{fig:buckboost01}
\end{figure}

The last of the basic converter topologies is the buck--boost converter. A circuit diagram is shown in  \Fref{fig:buckboost01}.
This converter consists of a DC input voltage source $V_\mathrm{S}$, a controlled switch $S$, an inductor $L$, a diode $D$, a filter capacitor $C$, and a load resistance $R$. While the switch $S$ is  `on', the inductor current increases and the diode is maintained `off'. When the switch is turned off, the diode provides a path for the inductor current. Note the polarity of the diode, which results in its current being drawn from the output. For this reason, the output voltage polarity is negative.

\begin{figure}
	\centering
	\includegraphics[scale=1]{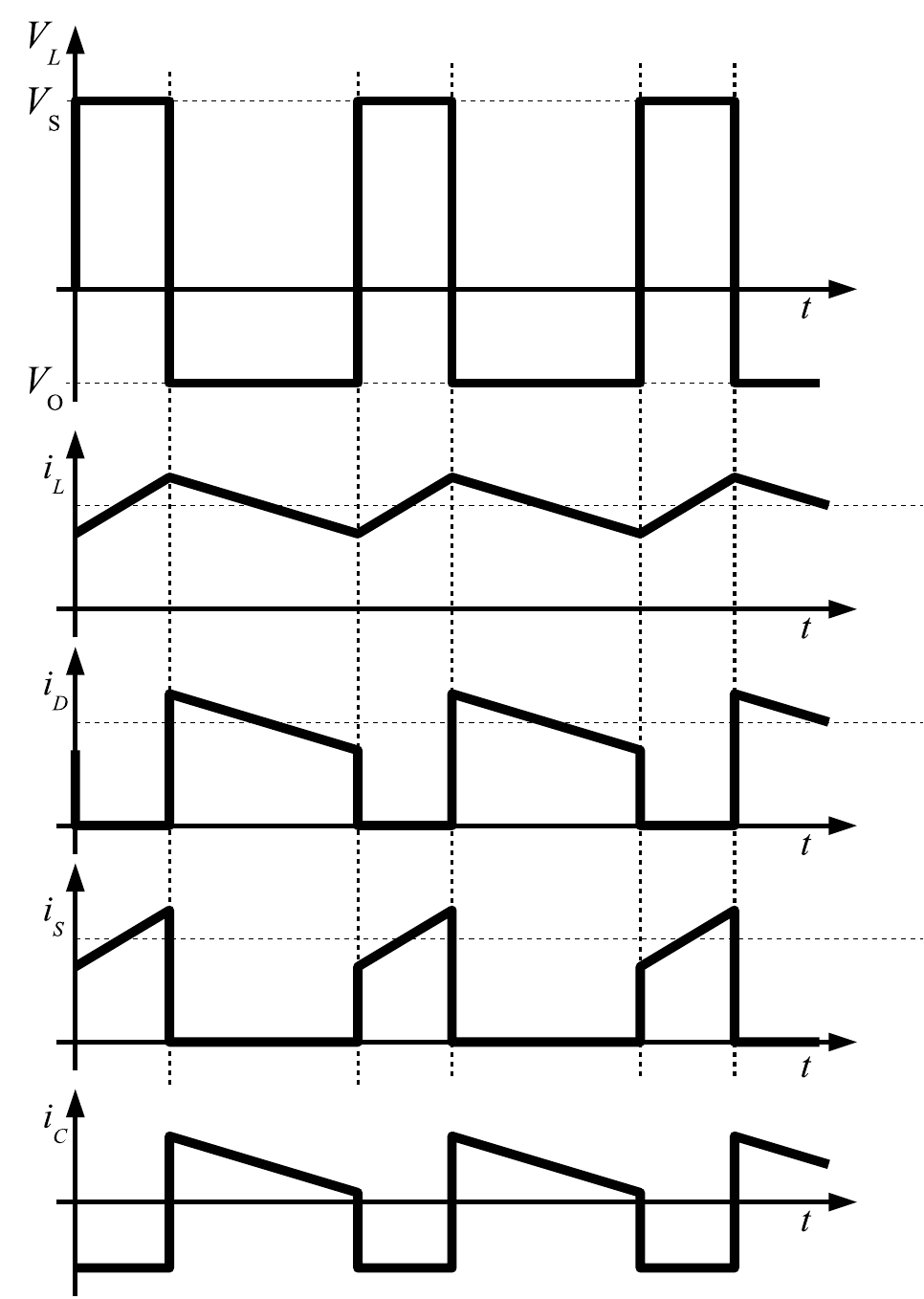}
	\caption{Waveforms for buck--boost converter}
	\label{fig:buckboost02}
\end{figure}

The waveforms for the buck--boost converter are depicted in  \Fref{fig:buckboost02}. The condition for a zero volt-second product for the inductor in the steady state  and the steady-state DC transfer function are given by the following equations:
\begin{equation}
V_\mathrm{S} T D = -V_\mathrm{O} (1 - D)T,
\label{eq:BB01}
\end{equation}
\begin{equation}
M_\mathrm{V}\equiv\frac{V_\mathrm{O}}{V_\mathrm{S}} = \frac{D}{(1-D)}.
\label{eq:BB02}
\end{equation}
The critical value of the inductor that fixes the boundary between the CCM and DCM is given by
\begin{equation}
L_\mathrm{b} = \frac{(1-D)^2 D R}{2 f}.
\label{eq:BB03}
\end{equation}

The current that feeds the RC output filter is the same as that for the boost converter but in the reverse direction. Hence the value of the capacitor is given by Eq. (\ref{eq:boost04}).

\section{Control of DC--DC converters}

One of the common features of the power converters presented in the previous section is that the output voltage depends on the duty cycle $D$, which was defined as the ratio between $t_\mathrm{on}$ and the total period of switching $T$. Changes in the duty cycle produce variations in the output voltage of the converter. In other words, the output voltage can be  controlled by changing the duty cycle.

The analysis of a converter from the point of view of control can be performed using the switching function.
The switching function is a mathematical tool to represent the state of a power converter. For the basic power converters described above, the switching function has two values. When the switch is on, the value of the switching function is unity, and when the switch is off, the value is zero:
 \begin{equation}
 s(t)= \left\{
 \begin{tabular}{cr}
 1 & when the switch is on,  \\
 0 & when the switch is off.  \\
 \end{tabular}
 \right.
 \label{eq:modulation01}
 \end{equation}

\begin{figure}
\centering
\includegraphics{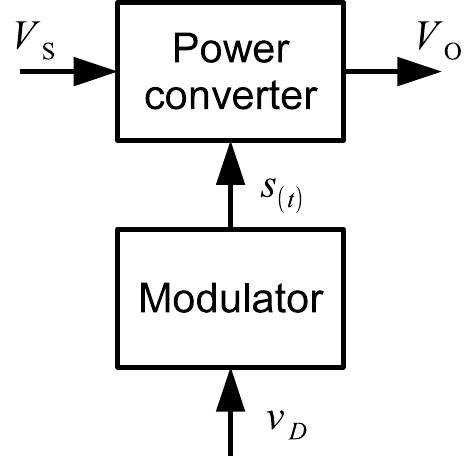}
\caption{Block diagram of power converter}
\label{fig:modulation01}
\end{figure}

A block diagram of the control of a power converter is shown in \Fref{fig:modulation01}. The block labelled ``Modulator'' transforms an input signal $v_D$ into the switching function $s(t)$, which controls the switch in the power converter. The modulator block can be built in several different ways. It can be implemented using a few analog components or using sophisticated digital circuits.

\begin{figure}
\centering
\includegraphics[width=0.9\linewidth]{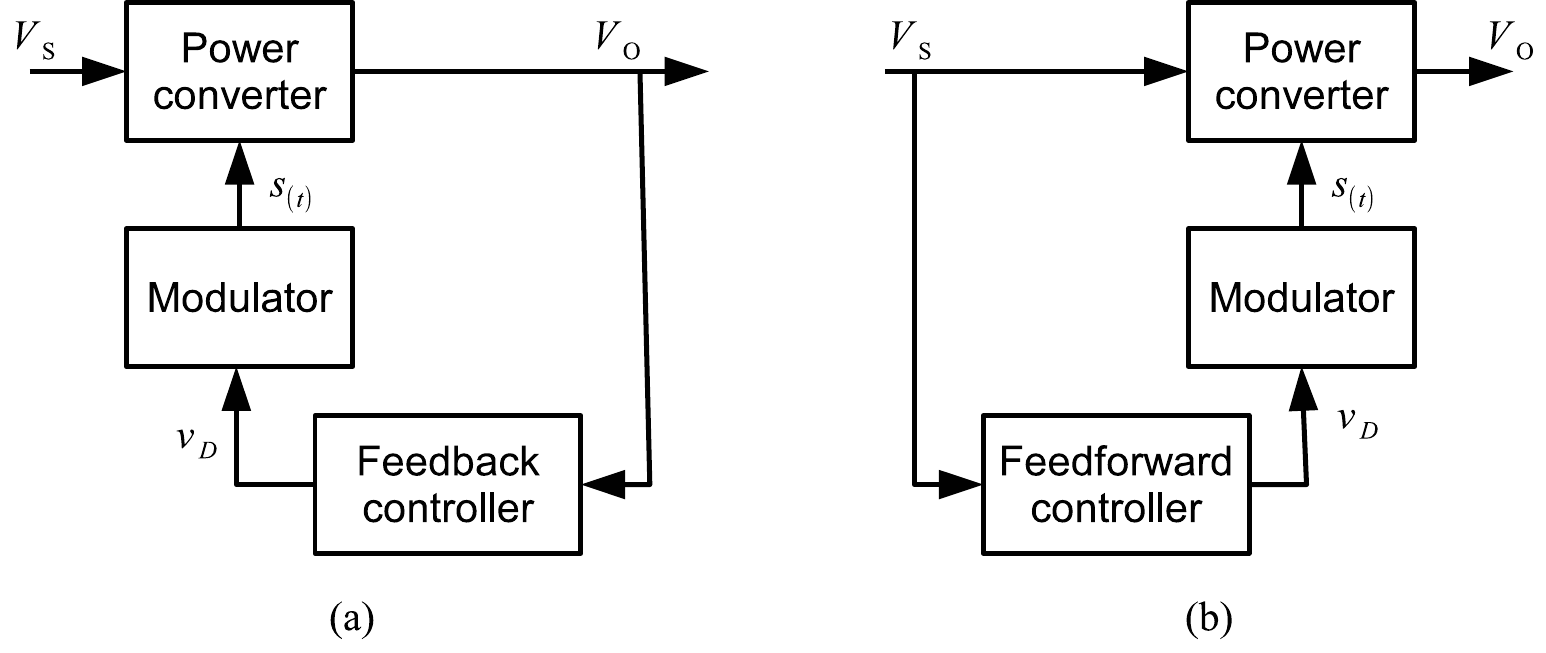}
\caption{Block diagrams of regulation schemes: (a) feedback regulation; (b) feedforward regulation}
\label{fig:modulation02}
\end{figure}

One of the most common applications of DC--DC converters is to provide a regulated output regardless of variations in the load and the input voltage. Another source of variation in the output voltage is changes in the components of the power converter; components can be affected by external factors such as temperature or ageing. The most common schemes used to regulate the output of power converters are presented in \Fref{fig:modulation02}.

The feedback scheme (\Fref{fig:modulation02}(a)) uses information from the output to provide the signal $v_D$ to the modulator. The main advantage of this scheme is that a feedback controller receives information directly from the output voltage, which is the value intended to be regulated, and therefore it can compensate for all variations regardless of their source. A disadvantage is that feedback regulation can lead to instability in the converter. Another disadvantage of the feedback scheme is that the output has to be affected in order to provide information to the feedback controller. In other words, the controller may not be fast enough to see a fluctuation and compensate for it.  For example, a variation in the input voltage $V_\mathrm{S}$ such as 100~Hz ripple from the rectification stage could have an amplitude such that the feedback scheme cannot completely reject it. In such cases a feedforward scheme, as shown in \Fref{fig:modulation02}(b), can be used. If it is known how the input voltage affects the output voltage (the DC transfer function), a feedforward controller can provide a signal $v_D$ to the modulator that gives the correct value of the output voltage.

The various ways of obtaining the switching function to be applied to the power converter are referred as control, or modulation, methods. A list and brief discussion of the most common methods are given in the following sections.

\subsection{Constant-frequency pulse width modulation (PWM)}
\begin{figure}
\centering
\includegraphics{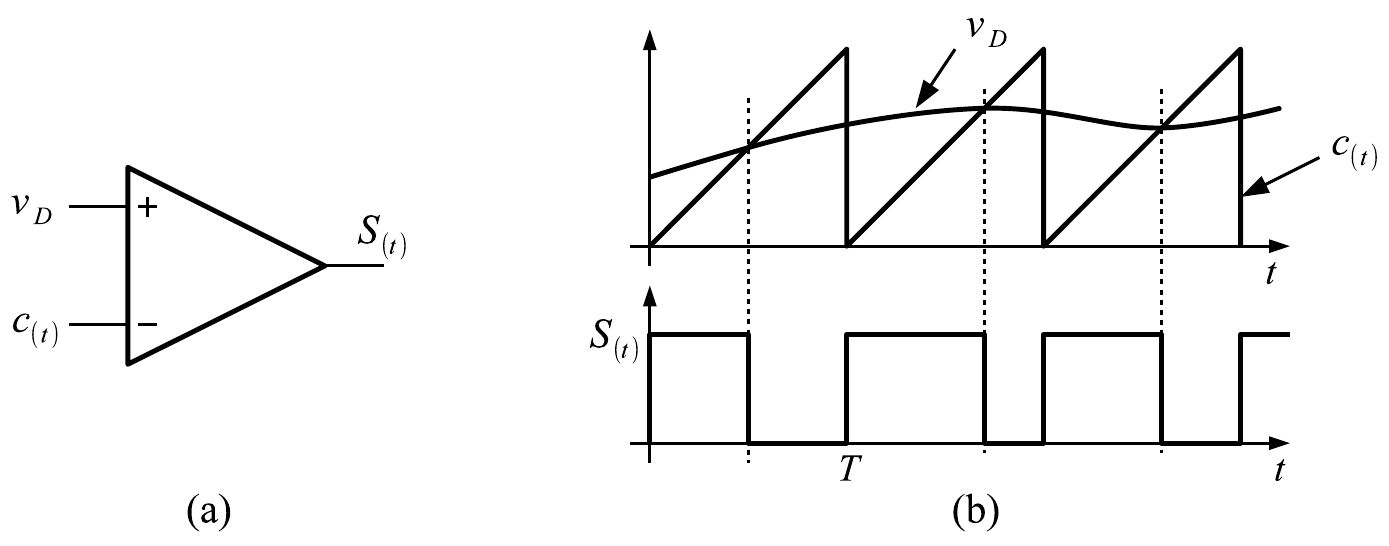}
\caption{PWM generation: (a) comparator; (b) waveforms}
\label{fig:pwm01}
\end{figure}

This is the most popular method of controlling power converters. In this modulation method, the period of the switching signal $T$ is constant and the information is contained in the width of the pulse, i.e., the period of time for which the switching signal is on, $t_\mathrm{on}$. This method is also known as carrier-based pulse width modulation. The switching function for constant-frequency PWM can be obtained by comparing the signal $v_D$ with a carrier signal $c(t)$ as depicted in \Fref{fig:pwm01}.

Figure \ref{fig:pwm01} shows an analog implementation of a PWM modulator. In digital implementations, the carrier is replaced by a counter/timer block. In both types of implementation, a way to avoid multiple transitions within a carrier period needs to be added.

\subsection{Variable-frequency PWM}
Variable-frequency PWM, although not as popular as constant-frequency PWM, is also used in practice. The common variations of variable-frequency PWM are the constant-off-time/variable-on-time and constant-on-time/variable-off-time versions.

\begin{figure}
\centering
\includegraphics{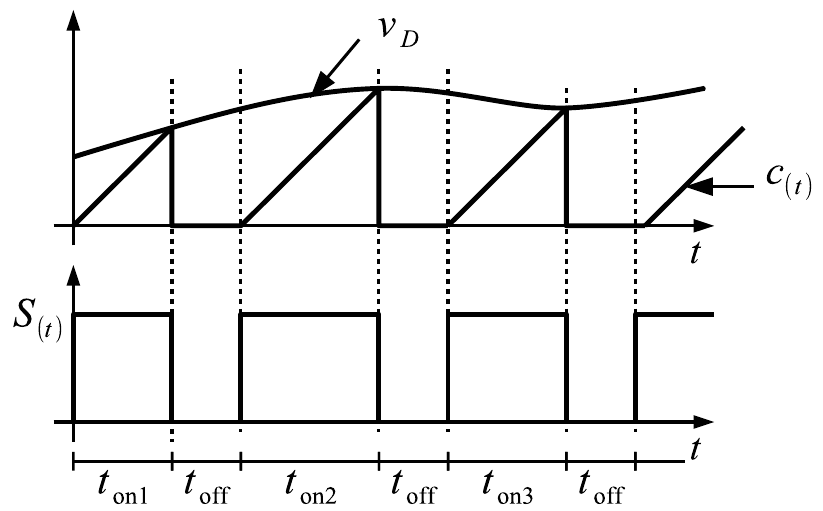}
\caption{PWM generation: constant-off-time/variable-on-time PWM}
\label{fig:pwm02}
\end{figure}

Figure \ref{fig:pwm02} shows the waveform for constant-off-time/variable-on-time PWM using a sawtooth-like carrier signal. The switch is turned on after a fixed time; at this point, the sawtooth signal starts to rise at a constant rate. The switch is turned off again when the sawtooth signal intersects the signal $v_D$.  After the turn-off of the switch, the carrier is reset to zero for a fixed period of time. The figure shows that the on time increases and the switching frequency decreases when the signal $v_D$ is increased, resulting in variable-frequency operation. Constant-on-time/variable-off-time PWM can be implemented in a similar way.
 	
Variable-frequency PWM is used in light-load operation of power converters in order to increase the efficiency of the converter under these conditions. The main disadvantage of this is the difficulty associated with the design of the input and output filters. The filter cut-off frequency has to be selected  based on the lowest possible switching frequency in order to provide the required attenuation of ripple and electromagnetic interference (EMI) under all possible operation conditions. This usually leads to a conservative design with significant volume and cost penalties.

\subsection{Current mode control}
\begin{figure}
\centering
\includegraphics[width=0.9\linewidth]{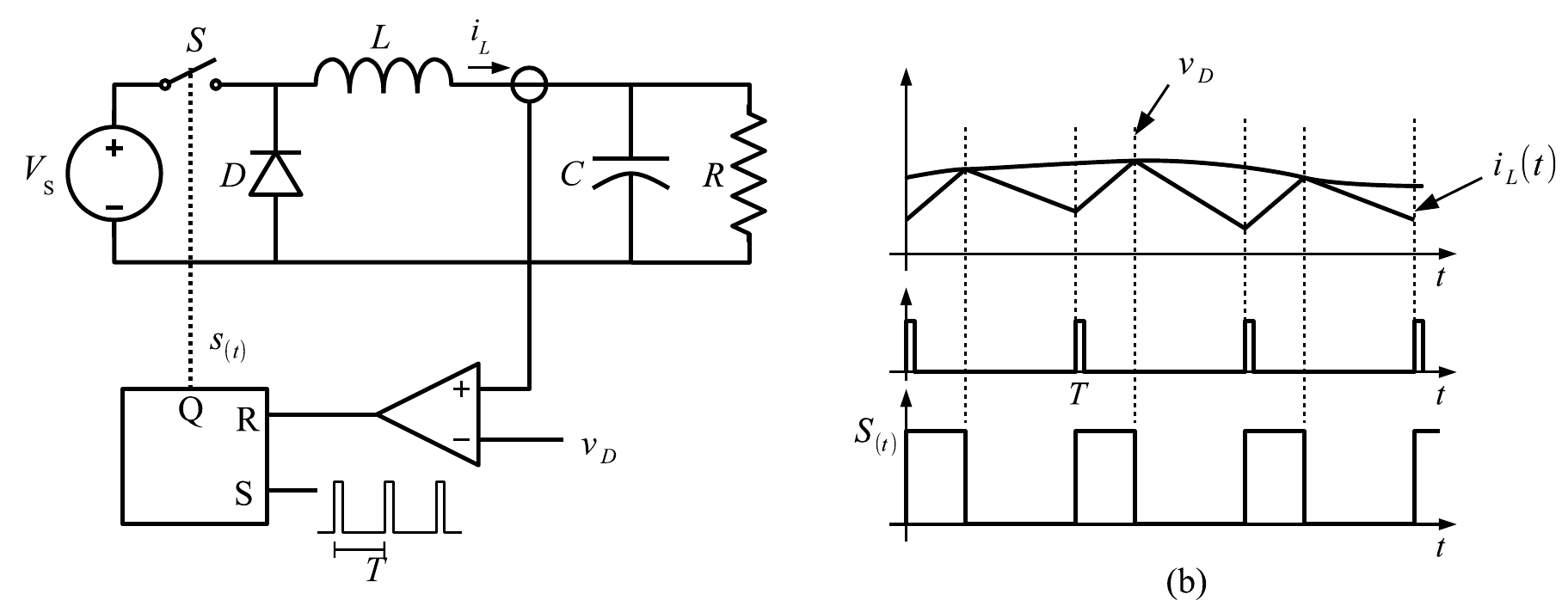}
\caption{Current mode: (a) circuit diagram; (b) waveforms}
\label{fig:CM01}
\end{figure}

Current mode control uses direct measurement of the inductor current to generate the switching function. One way of generating the switching function is to replace the sawtooth carrier waveform used to generate the PWM by the inductor current, as shown in \Fref{fig:CM01}(b). The switching function is set to unity by a clock signal of period $T$. When the inductor current reaches the value of the signal $v_D$, the switching function is reset. Figure \ref{fig:CM01}(a) shows a simple circuit for current mode PWM generation.

\subsubsection{Slope compensation} The peak current mode method is inherently unstable at duty cycles higher than 0.5, resulting in subharmonic oscillations. A compensating ramp (with slope equal to the inductor current down-slope) is usually applied to the comparator input to eliminate this instability.

Note that in current mode control, the output voltage is not directly controlled. An addition loop is needed in order to obtain a regulated output voltage.

\subsection{Variable-structure control}
Variable-structure control offers an alternative way to implement a switching function, which exploits the inherently variable nature of the structure of DC--DC converters. In practice, the converter switches are driven as a function of the instantaneous values of the voltages and currents. The most important feature of variable-structure control is its ability to provide robust control. Although the study of variable-structure control is far beyond on the scope of this article, hysteresis control provides a very simple and intuitive example. Sliding-mode control is other variable-structure control method for power converters reported in the literature.

\subsubsection{Hysteresis control}
\begin{figure}
\centering
\includegraphics[width=0.9\linewidth]{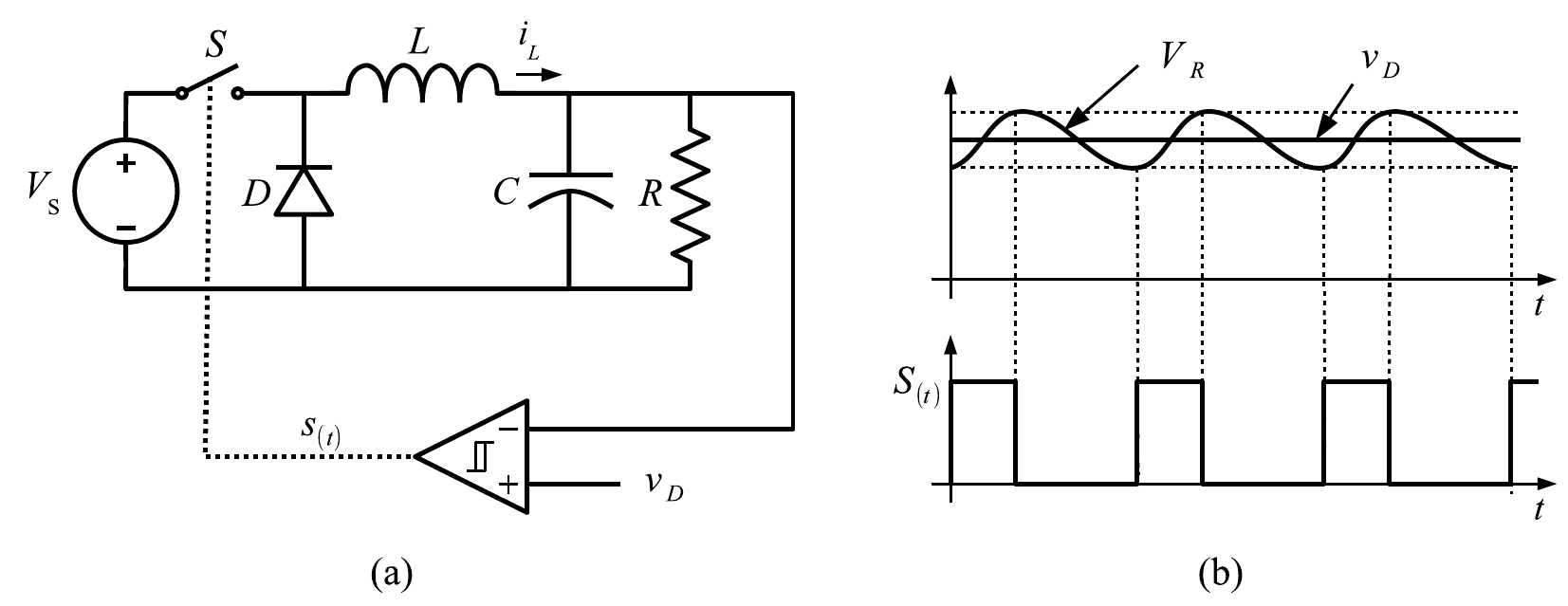}
\caption{Hysteresis control: (a) circuit diagram; (b) waveforms}
\label{fig:hyst01}
\end{figure}

Hysteresis control is based on comparing the output voltage $V_R$ with a reference voltage $v_D$ using a hysteresis comparator.
If the voltage is lower than the upper limit of the comparator, the switch is turned off until the output voltage reaches the lower limit of the hysteresis comparator. When the voltage reaches this limit, the switch is turned on. Figure \ref{fig:hyst01} shows a simple circuit diagram for hysteresis control, and typical waveforms.

\section{Derived converter topologies}\label{sec:derived}

Several different topologies have been derived from the basic DC--DC converters.  In this section,  three different groups of converters will be presented. The first group of converter topologies are isolated versions of the basic converters. The use of a transformer provides galvanic isolation between the input and output voltages. An additional advantage of the use of a transformer is an increase in the output voltage range of the converter due to the turns ratio of the transformer. The second group is formed by the flyback converter. This converter has galvanic isolation provided by coupled inductors. The difference between a transformer and coupled inductors is that in the latter case there is no direct energy flow from the input to the output. This fact limits the use of these converters to low-power applications. Finally, a converter with continuous input and output currents is presented.

\subsection{Half-bridge converter}
\begin{figure}
\centering
\includegraphics[width=0.9\linewidth]{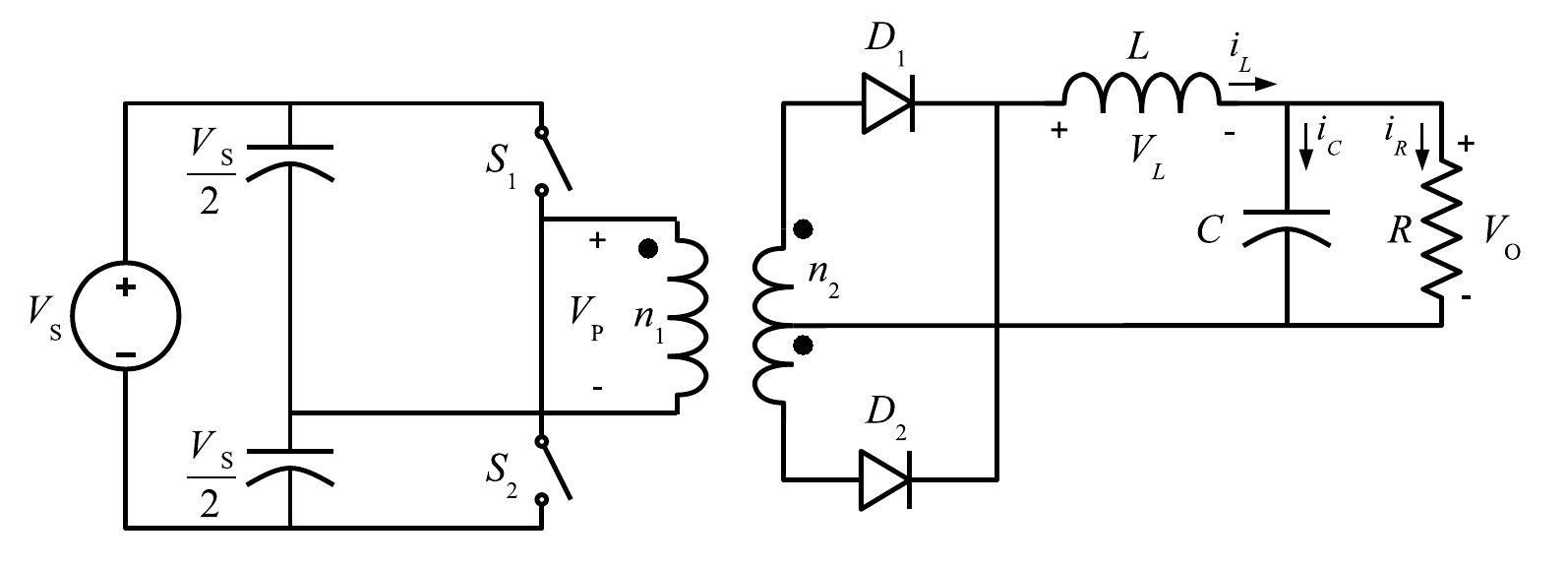}
\caption{Half-bridge converter: circuit diagram}
\label{fig:half01}
\end{figure}

\begin{figure}
\centering
\includegraphics{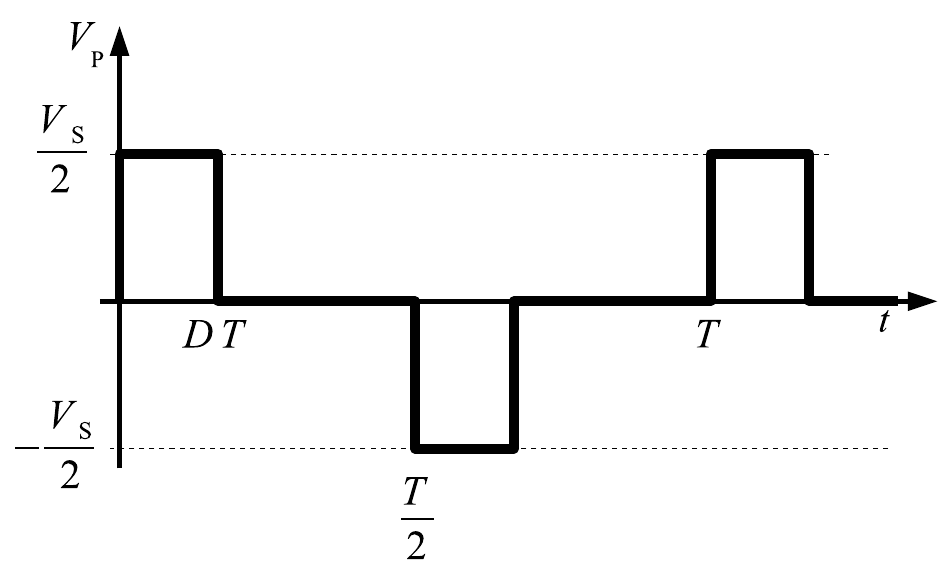}
\caption{Half-bridge converter: transformer primary voltage}
\label{fig:half02}
\end{figure}

The circuit diagram of a half-bridge converter is shown in \Fref{fig:half01}. The output filter part of the circuit is the same as that in a buck converter. The square waveform of the transformer secondary is rectified and then filtered by the output stage.
The voltage at the primary of the transformer is shown in \Fref{fig:half02}. The voltage per second
 on the primary and secondary sides during the switching period has to be zero in order to avoid saturation of the transformer.

In this converter, two capacitors act as an input voltage divider, which is connected to one of the inputs of the transformer. A two-polarity voltage is obtained by  connecting the other input to $V_\mathrm{S}$ or a reference voltage using the switches $S_1$ and $S_2$. The switches operate shifted in phase by $T$/2 with the same duty ratio $D$. The duty ratio must be smaller than 0.5. When the switch $S_1$ is on, the diode $D_1$ conducts and the diode $D_2$ is off. The diode states are reversed when the switch $S_2$ is on. When both controllable switches are off, both diodes are on and share the filter inductor current equally. This short-circuits the transformer and forces the transformer input voltage to zero. The DC voltage transfer function of the half-bridge converter is given by the following equation:
\begin{equation}
	M_\mathrm{V} \equiv \frac{V_\mathrm{O}}{V_\mathrm{S}} = 2 D \frac{n_2}{n_1}.
	\label{eq:half01}
\end{equation}

\subsection{Push--pull converter}
\begin{figure}
\centering
\includegraphics{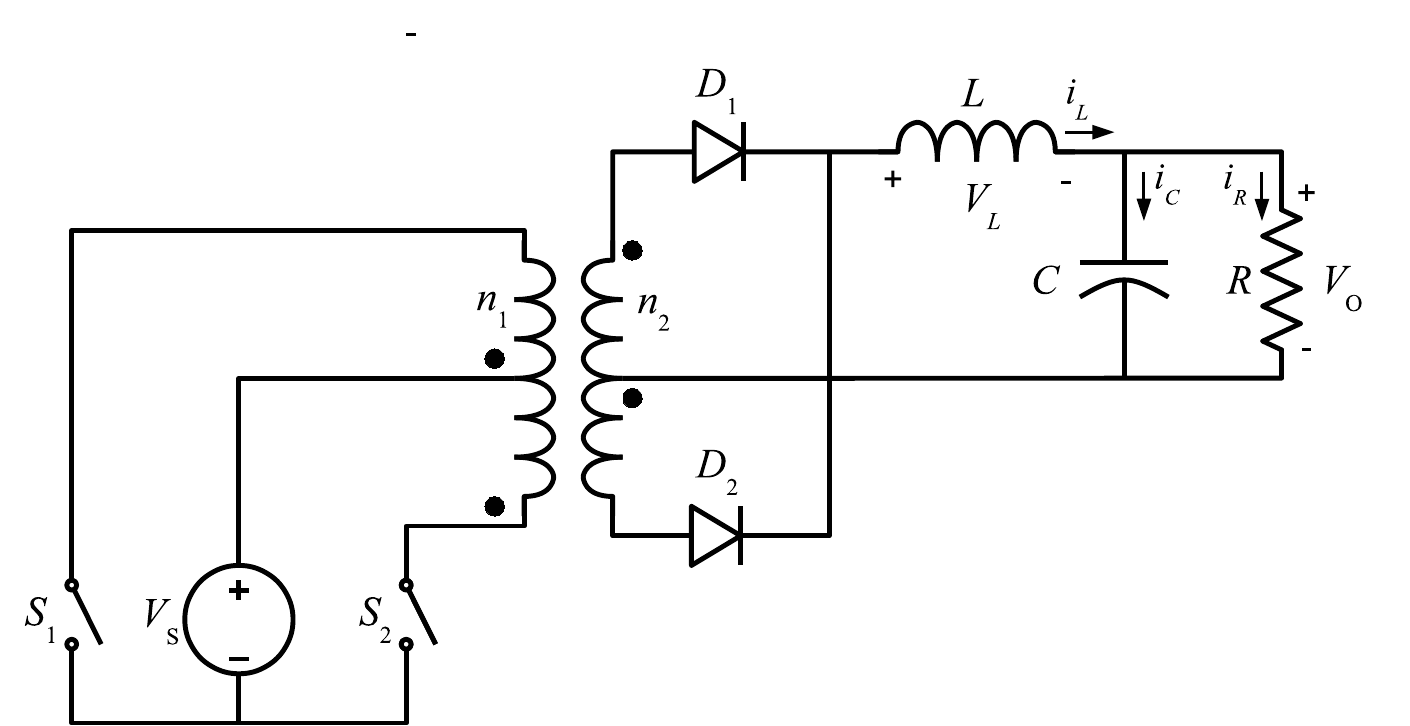}
\caption{Push--pull converter: circuit diagram}
\label{fig:push01}
\end{figure}

Push--pull converters, like half-bridge converters, can be regarded as two single-switch buck converters running out of phase. A circuit diagram is shown in \Fref{fig:push01}. The volts per second
 are balanced by alternating the operation of the switches $S_1$ and $S_2$. Similarly to the half-bridge converter, the output filter stage is the same as that in the buck converter. The main advantage of this topology is that both switches are connected to the reference voltage, which avoids the use of a floating driver for the switches. The main disadvantage is that differences in the period $t_\mathrm{on}$ of the switches could lead to saturation of the transformer.

The DC voltage transfer function of the push--pull converter is
\begin{equation}
M_\mathrm{V} \equiv \frac{V_\mathrm{O}}{V_\mathrm{S}} = 2 D \frac{n_2}{n_1}.
\label{eq:push01}
\end{equation}

\subsection{Full-bridge converter}
\begin{figure}[h]
\centering
\includegraphics[width=0.9\linewidth]
{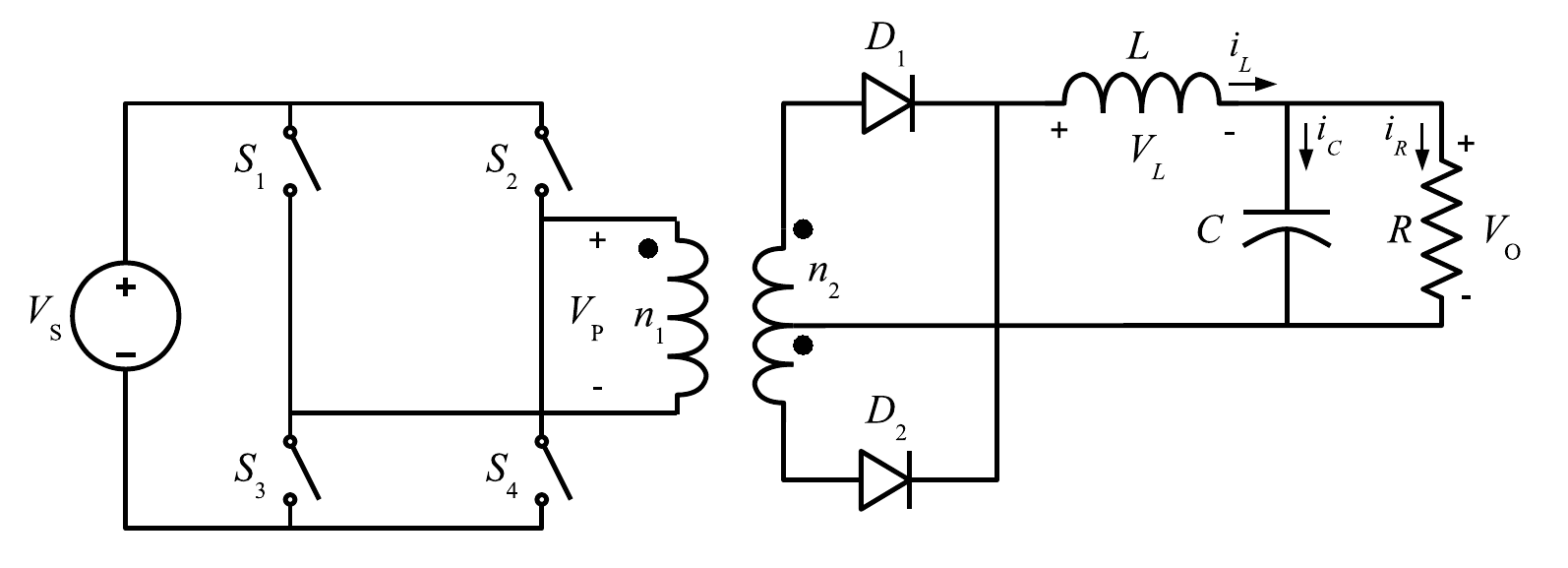}
\caption{Full-bridge converter: circuit diagram}
\label{fig:full01}
\end{figure}

The circuit diagram of a full-bridge converter is shown in \Fref{fig:full01}. In this topology, the controllable switches are operated in pairs. When $S_2$ and $S_3$ are on, a positive voltage $V_\mathrm{S}$ is applied to the primary side of the transformer. In this condition, the diode  $D_1$ conducts the inductor current. With  $S_1$ and $S_4$ on, the voltage applied to the transformer is $-V_\mathrm{S}$ and the the diode $D_2$  is on. If all the switches are off, both diodes conduct the output current and the secondary of the transformer is short-circuited.

The steady-state DC voltage transfer function is given by
\begin{equation}
M_\mathrm{V} \equiv \frac{V_\mathrm{O}}{V_\mathrm{S}} = 2 D \frac{n_2}{n_1}.
\label{eq:full01}
\end{equation}

\subsection{Forward converter}
\begin{figure}
\centering
\includegraphics{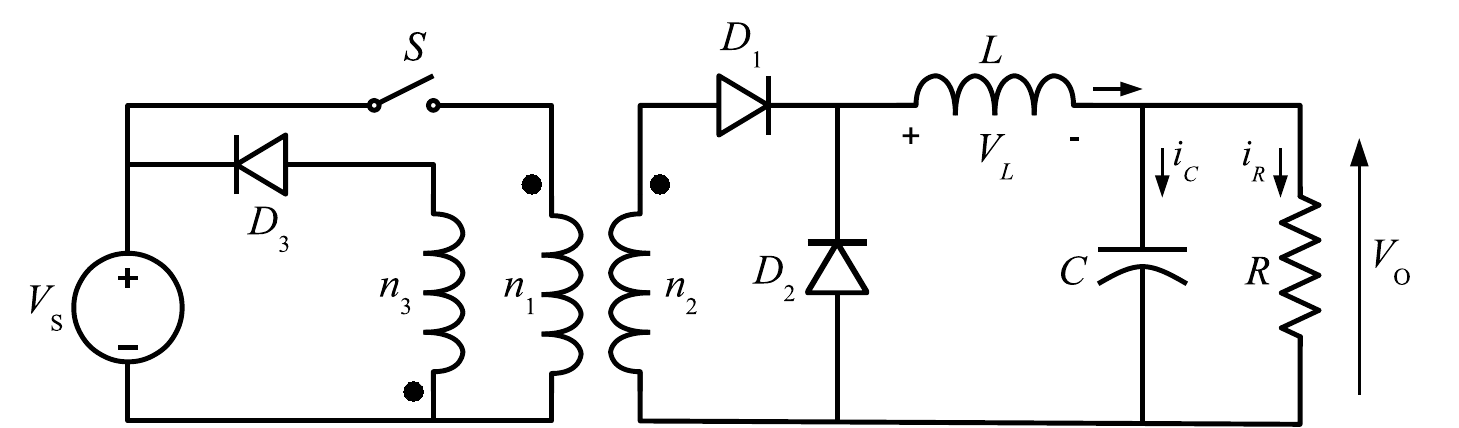}
\caption{Forward converter: circuit diagram}
\label{fig:forward01}
\end{figure}

The forward converter is another topology that uses a transformer in order to provide galvanic isolation. The  voltage output can be lower or higher than the input voltage depending on the transformer ratio. A circuit diagram is shown in \Fref{fig:forward01}. As in the previous isolated topologies, the output filter stage is identical to that in the buck converter. When the switch $S$ is closed, the diode $D_1$ conducts and transfers energy from the input to the output stage. During the period when the switch $S$ is open, the diode $D_2$ carries the current in the inductor $L$ and the diode $D_3$ connects the third transformer winding to the input voltage to decrease the magnetizing current to zero.

The steady-state DC voltage transfer function is given by
\begin{equation}
M_\mathrm{V}\equiv\frac{V_\mathrm{O}}{V_\mathrm{S}} = 2 D \frac{n_2}{n_1}.
\label{eq:forward02}
\end{equation}
The condition to avoid saturation of the transformer is
\begin{equation}
n_1 D \leq n_3 (1-D).
\label{eq:forward01}
\end{equation}

\subsection{Flyback converter}
\begin{figure}
\centering
\includegraphics{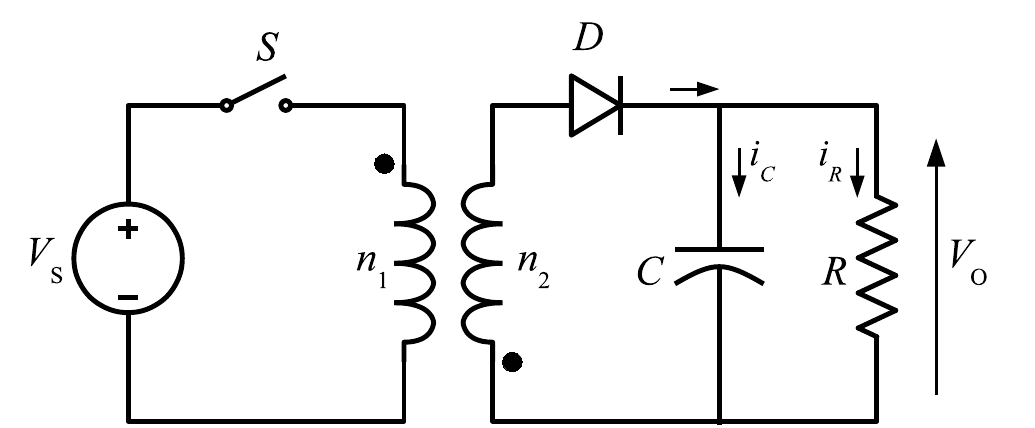}
\caption{Flyback converter: circuit diagram}
\label{fig:flyback01}
\end{figure}

The flyback converter is derived from the buck--boost topology. The inductor is split into two coupled inductors. As a result of this, galvanic isolation is achieved and the DC voltage transfer function is multiplied by the turns ratio of the inductors. The basic circuit diagram is shown in \Fref{fig:flyback01}.

The magnetic components of the flyback converter act more like two inductors sharing a common magnetic core than as a transformer. Energy is stored in the common magnetic core during the time $t_\mathrm{on}$ of the switch $S$ and is then transferred to the output stage during the time $t_\mathrm{off}$. There is no direct energy transfer between the primary and secondary windings, unlike the converters mentioned above. This limits the application of this converter to low-power applications.

\subsection{\'{C}uk converter}
\begin{figure}
	\centering
	\includegraphics{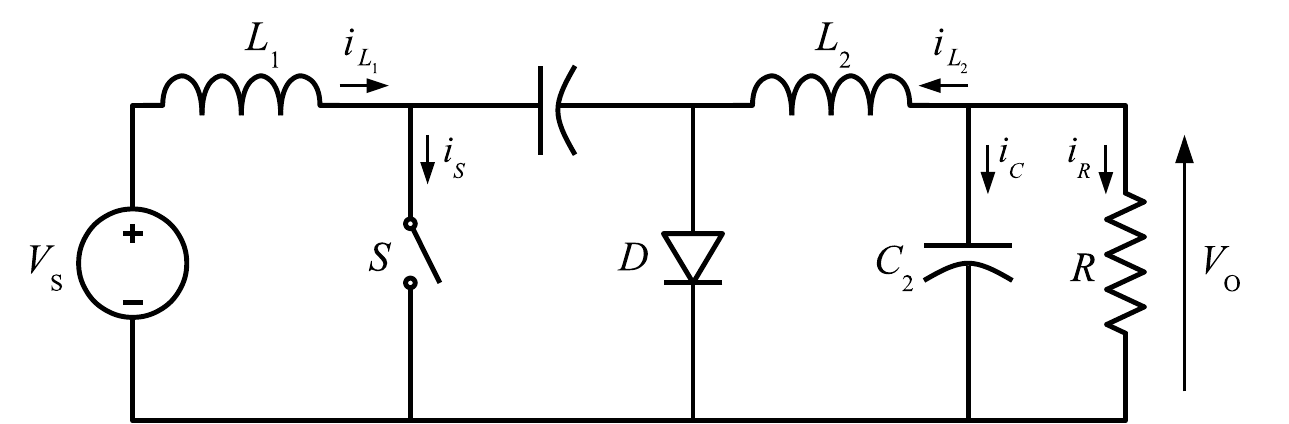}
	\caption{\'{C}uk converter: circuit diagram}
	\label{fig:cuk01}
\end{figure}

The \'{C}uk (pronounced `chook') converter was introduced by Slobodan \'{C}uk of the California Institute of Technology. The circuit diagram of this converter is shown in \Fref{fig:cuk01}. The main advantage of this converter is the continuous currents at the input and output of the converter. The main disadvantage is the high current stress on the switch.

The DC voltage transfer function of the converter is obtained using the average current in the capacitor $C_1$, which should be zero during a switching period when there is power balance between the input and output of the converter:
\begin{equation}
I_{L_2} D T = I_{L_1} (1-D) T ,
\label{eq:cuk01}
\end{equation}
\begin{equation}
P_\mathrm{IN} = V_\mathrm{S} I_{L_1} =  -V_\mathrm{O} I_{L_2} = P_\mathrm{OUT}.
\label{eq:cuk02}
\end{equation}
Using these two equations,
\begin{equation}
M_\mathrm{V} \equiv \frac{V_\mathrm{O}}{V_\mathrm{S}} = \frac{D}{(1-D)}.
\label{eq:cuk03}
\end{equation}

Isolated versions of the \'{C}uk converter can be found in the literature. There are also versions of this converter in which the two inductors $L_1$ and $L_2$ are integrated into one magnetic element in order to reduce the cost and volume of the converter.

The list of converters presented in this section is not intended to be exhaustive. The converter topologies presented are just a small sample of the many topologies that have been reported in the literature. The right topology for each application depends on many factors, which include the stress on the semiconductors and passive components, the level of power to be transferred, and the switching frequency, among others.

\section{Additional topics}\label{sec:topics}

This section presents some additional topics related to one-quadrant power converters that are worth mentioning. Firstly, synchronous rectification is addressed as a method to reduce losses, especially in low-voltage power converters. Interleaved converters are discussed as a way to increase the power handled by converters while reducing the stress on the input and output filters. Finally, a brief introduction to hard switching, snubbers, and  soft switching is given.

\subsection{Synchronous rectification}
One component of most one-quadrant converter topologies is a diode. This diode is used to provide a path for the current when the main switch is turned off, or for rectification and providing a DC output. The conduction losses of the diode make an important contribution to the overall power loss of a converter, especially for low-output-voltage, high-output-current power converters.

The conduction loss of a diode in a buck converter is given by the following equation:
\begin{equation}
P_\mathrm{D} = V_\mathrm{F} I_\mathrm{O} (1-D).
\label{eq:synch01}
\end{equation}
The overall efficiency due to the diode conduction loss can be expressed as
\begin{equation}
\nu  = \frac{V_\mathrm{O} I_\mathrm{O}}{V_\mathrm{O} I_\mathrm{O} + V_\mathrm{F} I_\mathrm{O} (1-D)} = \frac{V_\mathrm{O}}{V_\mathrm{O} + V_\mathrm{F} (1-D)}.
\label{eq:synch02}
\end{equation}

\begin{figure}
	\centering
	\includegraphics{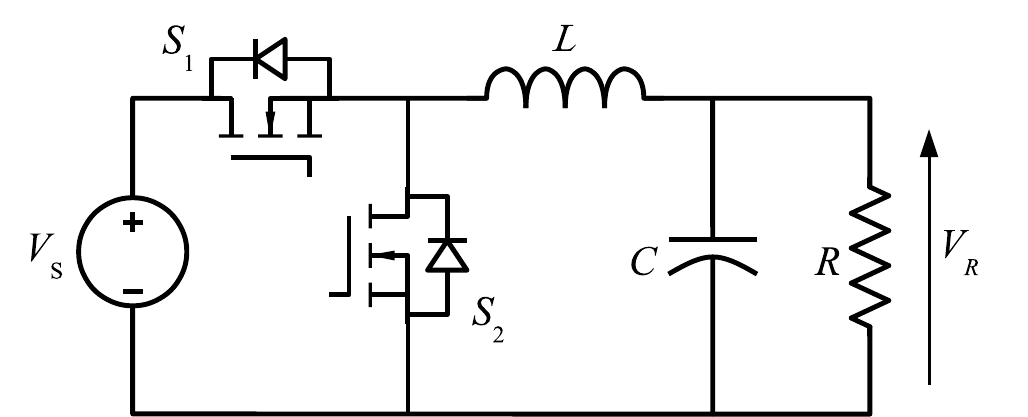}
	\caption{Synchronous rectification: circuit diagram}
	\label{fig:synch01}
\end{figure}

\begin{figure}
	\centering
	\includegraphics[scale=0.85]{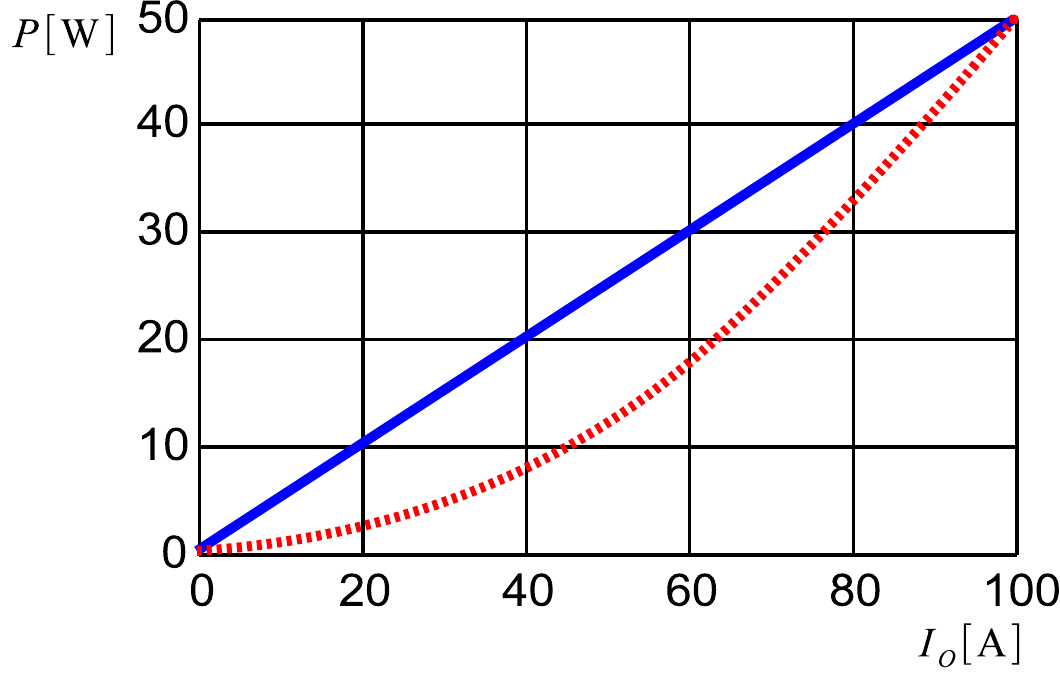}
	\caption{Power loss in $S_2$: diode (solid line) and MOSFET (dashed line)}
	\label{fig:synch02}
\end{figure}

For low-voltage power converters, this equation shows that the overall efficiency of the converter can drop to unacceptable levels. One way to increase the overall efficiency under these conditions is to replace the diode by a MOSFET, as shown in \Fref{fig:synch01}. The MOSFET introduces an almost linear resistance with a lower voltage drop. Figure \ref{fig:synch02} shows the power loss in $S_2$ for a diode and for a MOSFET.

In addition to minimizing conduction losses, MOSFETs offer the additional benefit that they can easily be paralleled because their $R_\mathrm{DS_{on}}$ has a positive temperature coefficient.

\subsection{Interleaved converters}
\begin{figure}[]
\centering
\includegraphics[width=0.9\linewidth]{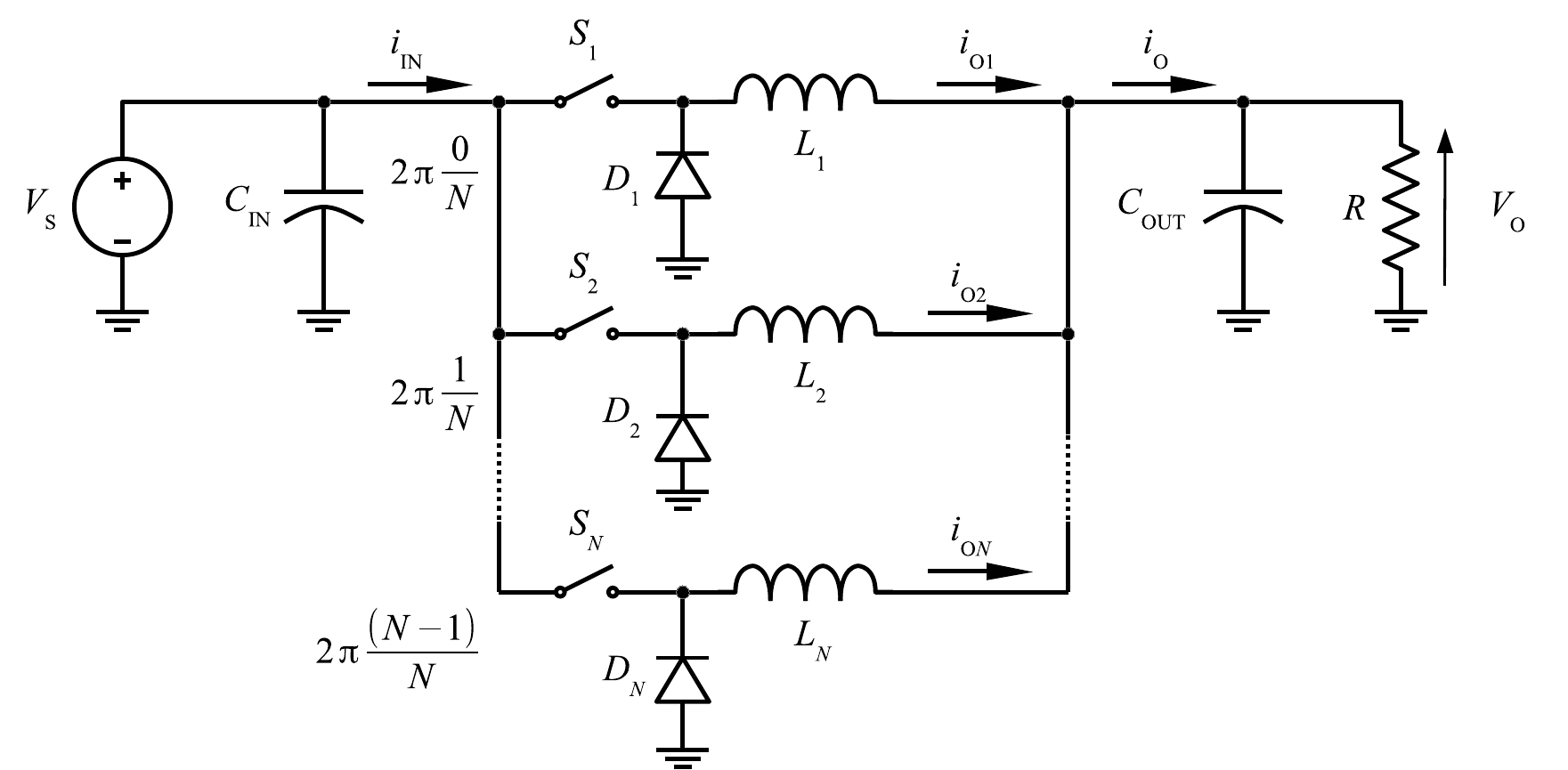}
\caption{Interleaved buck converter}
\label{fig:inter01}
\end{figure}

The concept of paralleling several DC--DC power converters in order to increase the output power is well known in power electronics. A further step is to group the components together and control the resulting converter as a unit. Interleaving is one of the techniques most often used, and provides several advantages such as an increase in the resulting switching frequency and a reduction in the component size. Figure \ref{fig:inter01} shows the circuit diagram of an interleaved buck converter. The switching functions for the switches $S_1, S_2, \ldots , S_N$ have the same switching frequency but there is a phase shift between them equal to $2\pi/N$, where $N$ is the number of individual stages that form the converter.

\begin{figure}
\centering
\includegraphics[scale=0.6]{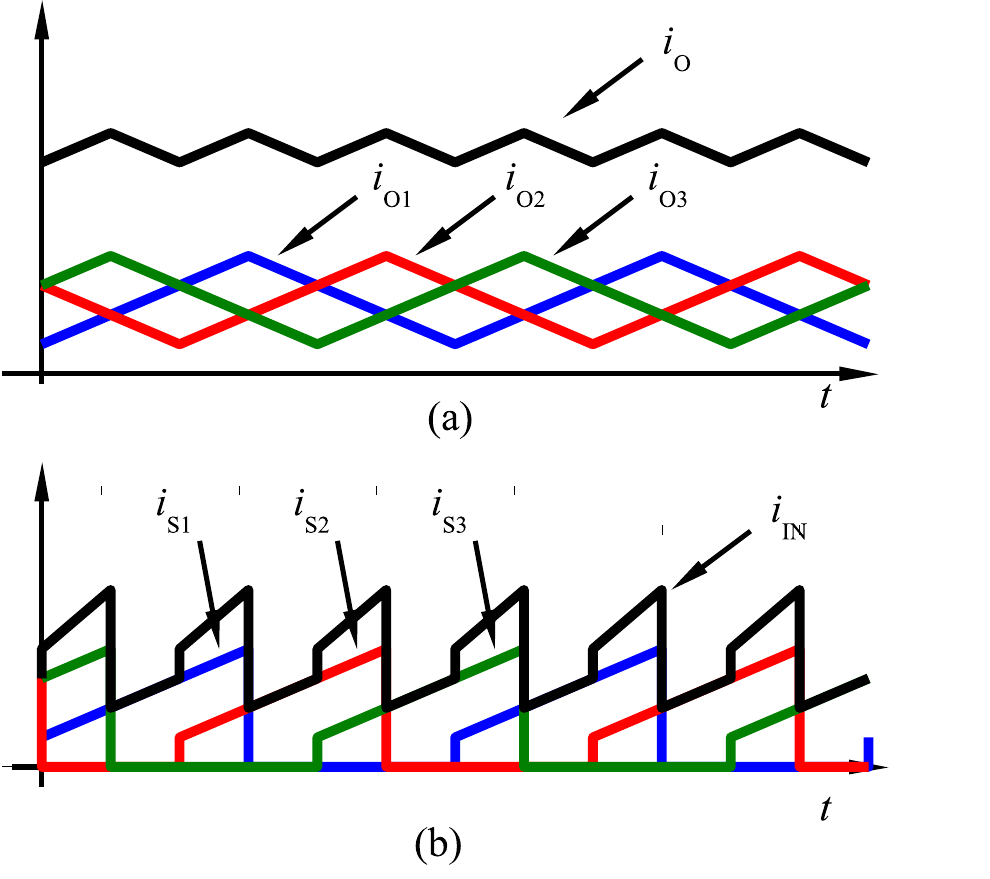}
\caption{Output and input currents for a three-stage interleaved buck converter}
\label{fig:inter02}
\end{figure}

\begin{figure}
\centering
\includegraphics[scale=0.5]{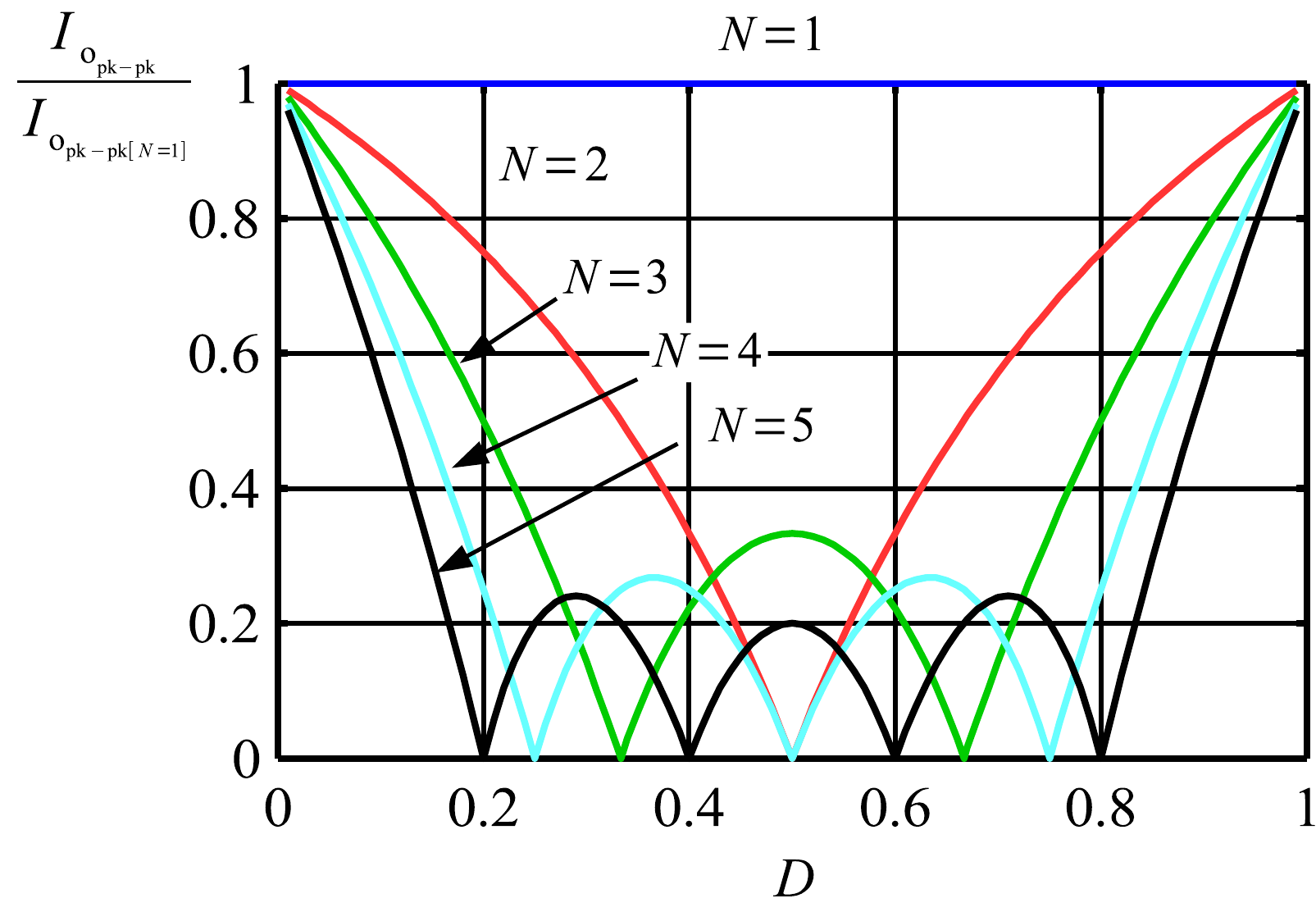}
\caption{Normalized output ripple current}
\label{fig:inter03}
\end{figure}

The first plot in \Fref{fig:inter02} shows the output currents of the individual stages and the overall output current formed by addition of the inductor currents for each phase. The ripple frequency of the overall output current is $N$ times the frequency of the inductor phase currents. This multiplication of the ripple frequency reduces the requirements on the output filter. Figure \ref{fig:inter03} shows the normalized output ripple current. This shows how the ripple is reduced; for particular values of $D$, the ripple current is zero.

\begin{figure}
\centering
\includegraphics[scale=0.55]{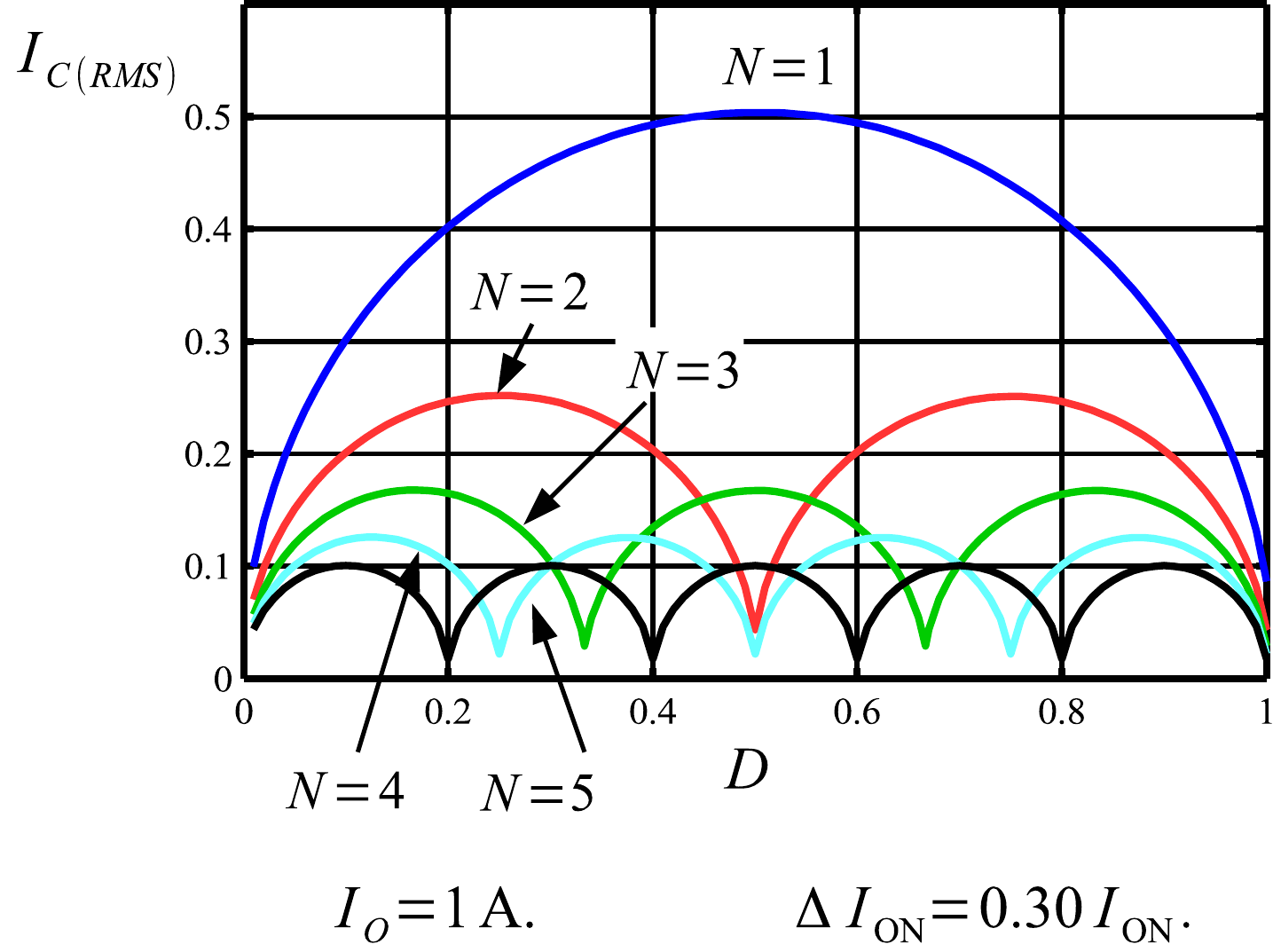}
\caption{Normalized input capacitor RMS current}
\label{fig:inter04}
\end{figure}

A similar effect is produced in the input current. The requirements on the input capacitor are also reduced by the interleaving of the converters. Figure \ref{fig:inter02}(b) shows the input current for each branch and the total input current. An increase in the frequency and a reduction in the RMS value can be observed. Figure \ref{fig:inter04} shows the normalized input capacitor RMS current for different values of $N$.

\subsection{Hard switching, snubbers, and soft switching}
\begin{figure}
\centering
\includegraphics[width=0.8\linewidth]{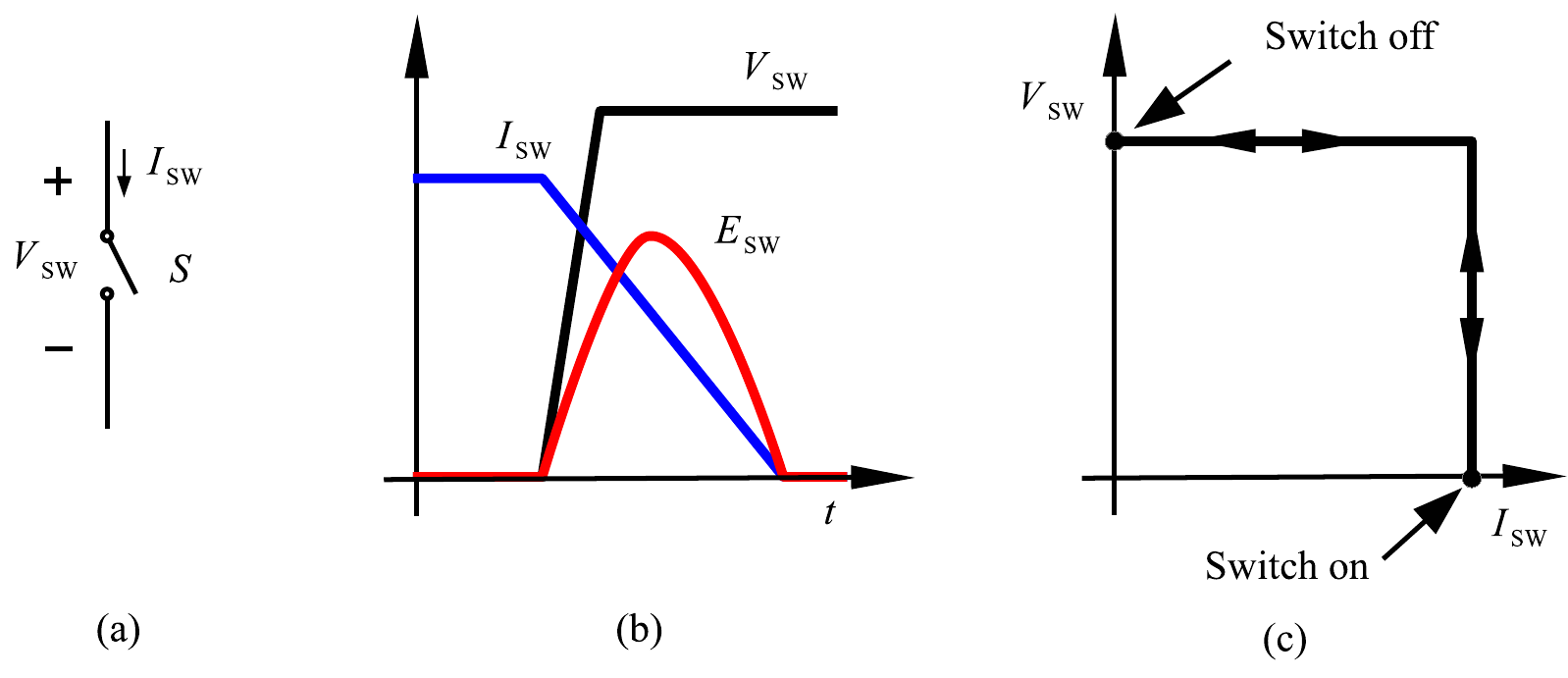}
\caption{Hard switching: (a) single switch without snubber; (b) switch voltage and current during transition; (c) transition diagram.}
\label{fig:hardSW}
\end{figure}

One of the big advantages of operating a power converter in switched mode is that voltages across and currents through the switches are not present simultaneously during the `on' and `off' states. This cannot be true, however, during the transitions between the states of the converter. The trajectories during the transitions determine whether the switches operate in a hard or soft switching mode.
The switching trajectory of a hard-switched power device is shown in \Fref{fig:hardSW}. During the turn-on and turn-off processes, the power device has to withstand high voltages and currents simultaneously, resulting in high switching losses and stress.

The switching losses are proportional to the operating frequency of the converter and are usually the main factor that limits the operating frequency. Another disadvantage of hard-switching operation is an increase in high-frequency emissions, which produce EMI.

\begin{figure}
\centering
\includegraphics[width=0.9\linewidth]{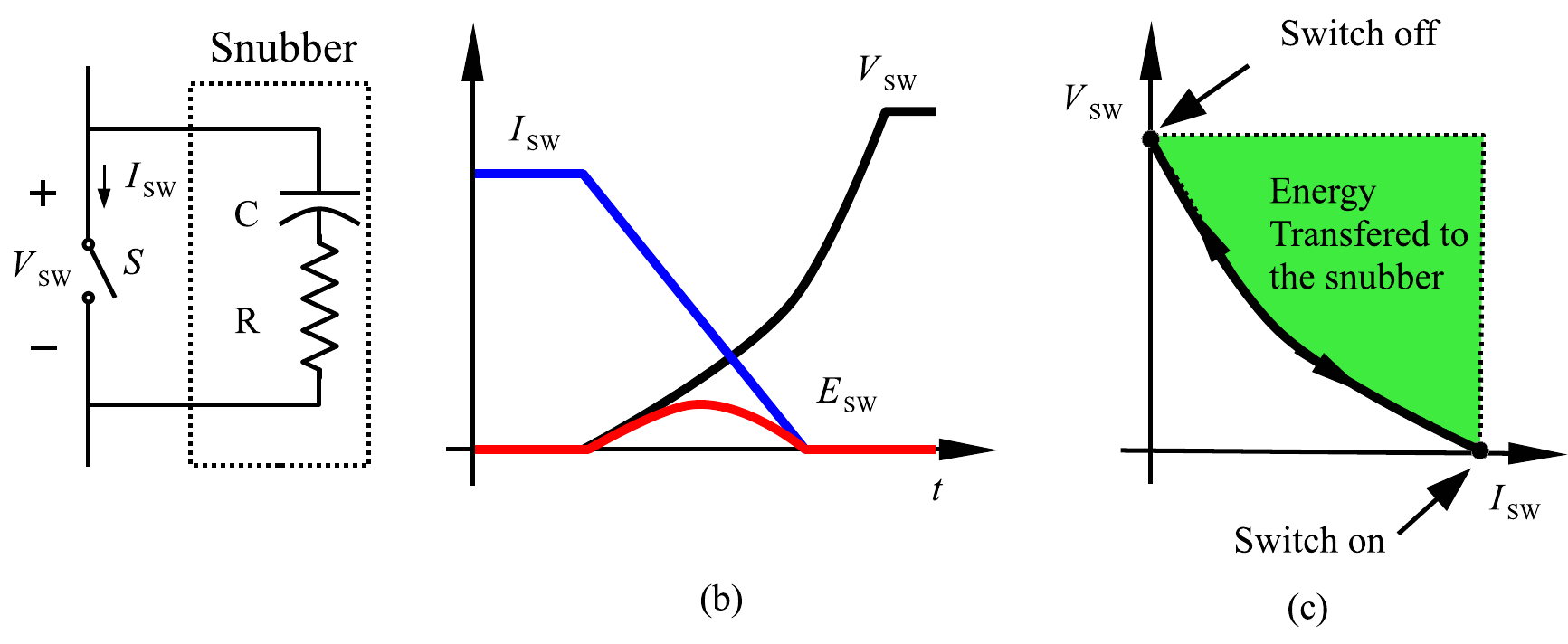}
\caption{Snubbers: (a) circuit diagram of RC snubber; (b) switch voltage and current during transition; (c) transition diagram.}
\label{fig:snubbers}
\end{figure}

Dissipative passive snubbers are usually added to  power circuits in order to reduce the $\mathrm{d}v/\mathrm{d}t$ and $\mathrm{d}i/\mathrm{d}t$ of the power devices, and the switching losses and stresses can be diverted to these passive snubber circuits. Figure \ref{fig:snubbers} shows how the transition trajectory is modified by the presence of an RC snubber. The $\mathrm{d}v/\mathrm{d}t$ of the switch is reduced during the turn-off and part of the transition energy is transferred to the dissipative element of the snubber.

The snubber design has to be done carefully in order to correctly dimension the components and ensure proper operation. More complex snubber circuits have the ability to recover the transition energy and increase the overall efficiency of the converter.

\begin{figure}
	\centering
	\includegraphics[scale=0.8]{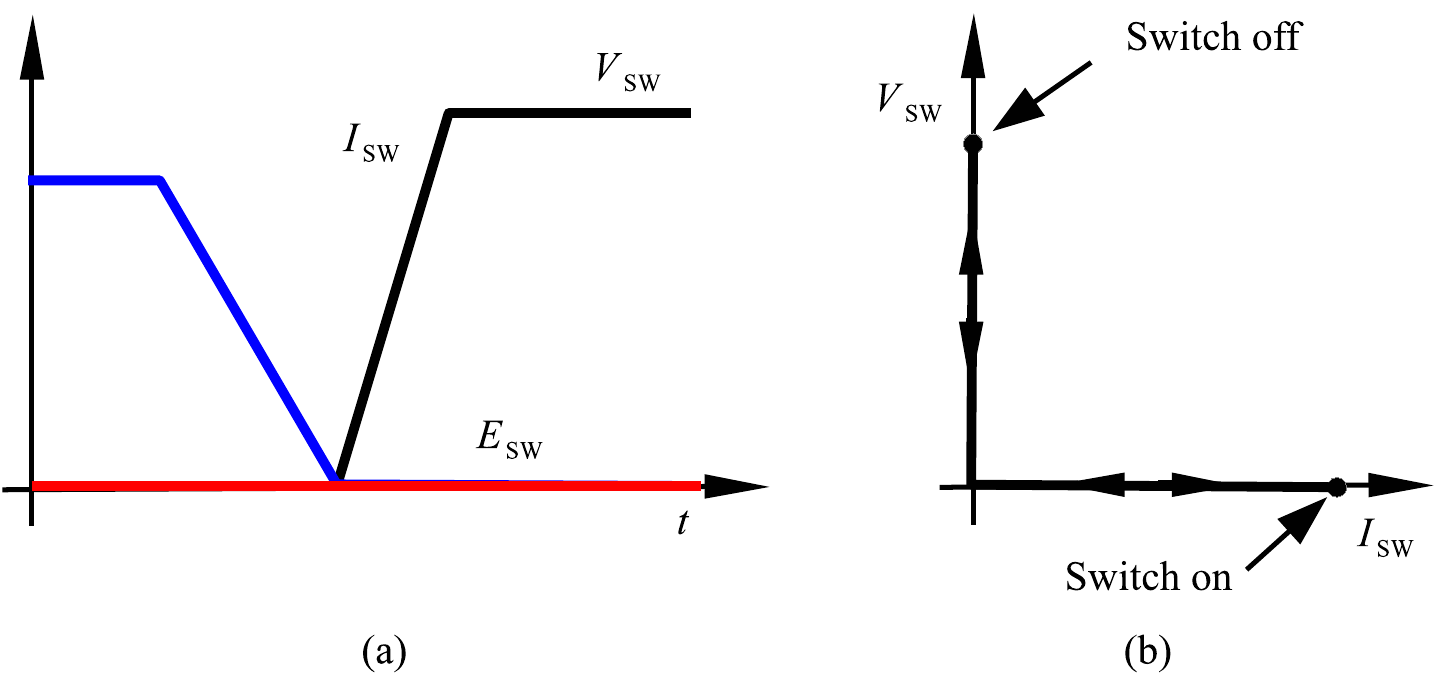}
	\caption{Soft switching: (a) switch voltage and current during transition; (b) transition diagram}
	\label{fig:SoftSwitching01}
\end{figure}

A further step can be taken to reduce the commutation losses by operating the switches in a soft switching mode. The basic idea of soft switching can be explained using the plots in \Fref{fig:SoftSwitching01}. The voltage across the switch is forced to be zero while the current is dropping. When the current is zero, the switch voltage starts to rise until the switch reaches the turn-off state. The losses during the transition are zero.

\begin{figure}
	\centering
	\includegraphics[width=0.75\linewidth]{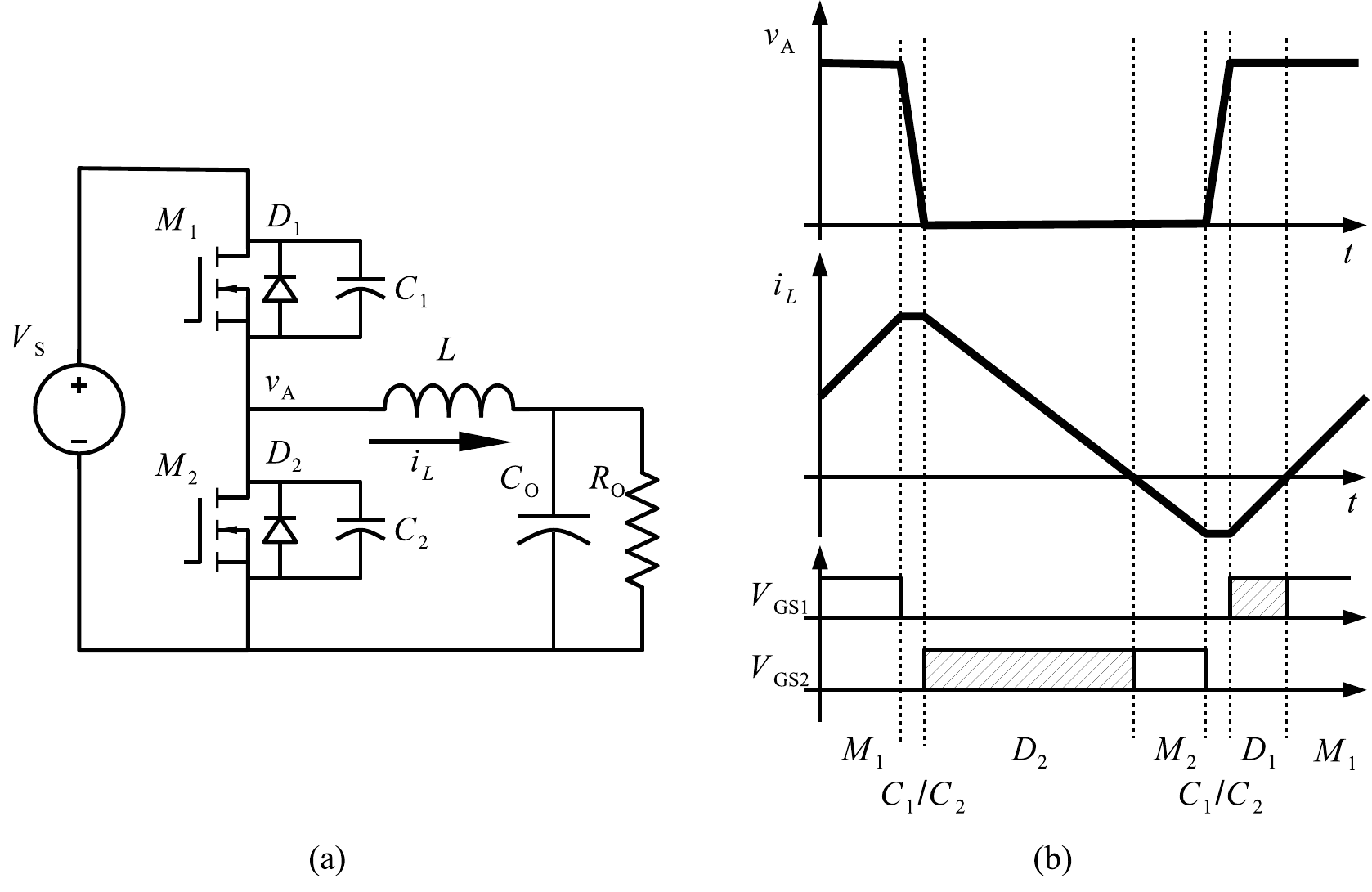}
	\caption{Soft switching: (a) circuit diagram; (b) waveforms}
	\label{fig:SoftSwitching02}
\end{figure}

A simple circuit operating in a soft switching mode is shown in \Fref{fig:SoftSwitching02}. The switch $M_1$ opens with zero voltage across its terminals because this voltage is maintained by the capacitor $C_1$. The voltage $v_\mathrm{A}$ drops because the inductor current charges and discharges the capacitors $C_1$ and $C_2$. Once the voltage reaches zero, the diode $D_2$ starts to carry current and k-eps-converted-to.pdf the voltage close to zero. During the period of time that the diode is conducting, the switch $M_2$ is turned on. The positive transition of voltage $v_\mathrm{A}$ is produced in a similar way.

\section{DC--DC resonant converters}

The trend in power electronics is towards an increase in efficiency, power, and component density. For this reason, resonant converters have received special interest in recent years. These converters have the potential to operate at higher frequencies and with lower switching losses than hard-switching converters.
The design of resonant converters and their control presents challenges different from those for other converters. In particular, the control of these converters is usually done by frequency modulation instead of pulse width modulation. Although a comprehensive analysis of DC resonant converters is far beyond the scope of this article, a brief description of their functional principles is given here.

All of the resonant-converter topologies operate in essentially the same way. The power switches generate a square wave voltage or current, which is applied to a resonant circuit. Energy circulates in the resonant circuit, and part of this energy is transferred to the output.

\begin{figure}
	\centering
	\includegraphics[width=0.9\linewidth]{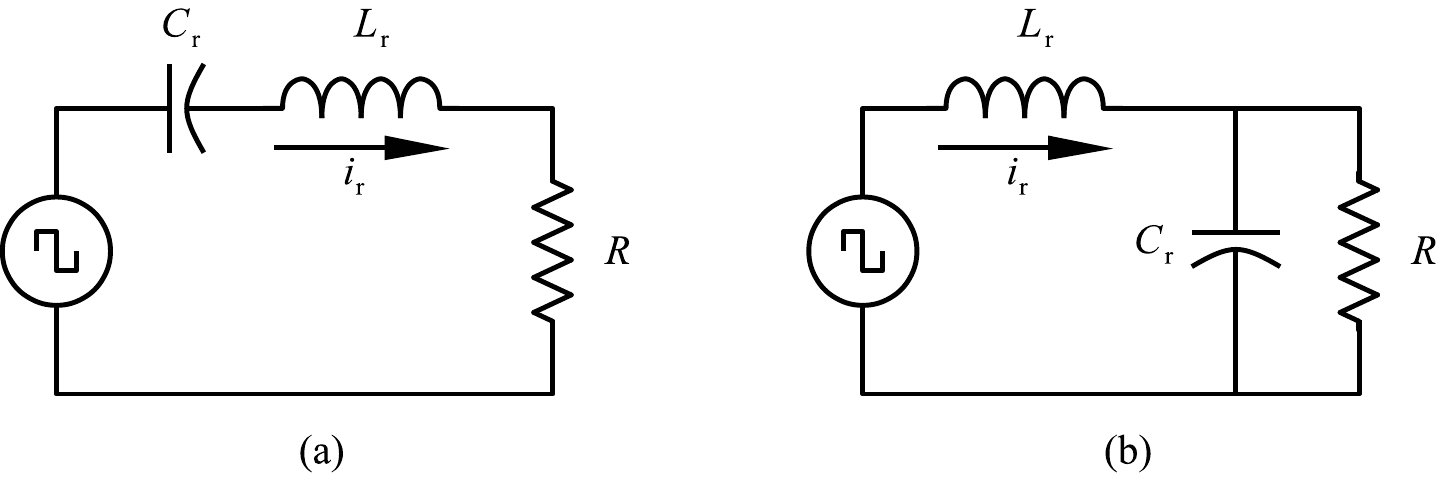}
	\caption{(a) Series resonant converter; (b) parallel resonant converter}
	\label{fig:Resonant01}
\end{figure}

The two basic types of resonant converter are the series resonant converter, shown in \Fref{fig:Resonant01}(a), and the parallel resonant converter, shown in \Fref{fig:Resonant01}(b). In both topologies, the energy transferred to the output is regulated by changing the frequency of the driving voltage. The resonant circuit forms a voltage divider with the resistor $R$. By changing the frequency, the impedance of the resonant circuit and therefore the output voltage can be varied.

The circuits shown in \Fref{fig:Resonant01} have limitations. The series resonant converter cannot work under very light load conditions, because the operating frequency would have to be very high in order to regulate the output voltage. The parallel resonant converter requires large amounts of circulating current when operating under heavy load conditions.

\subsection{LCC and LLC resonant converters}
\begin{figure}
	\centering	
	\includegraphics[width=0.9\linewidth]{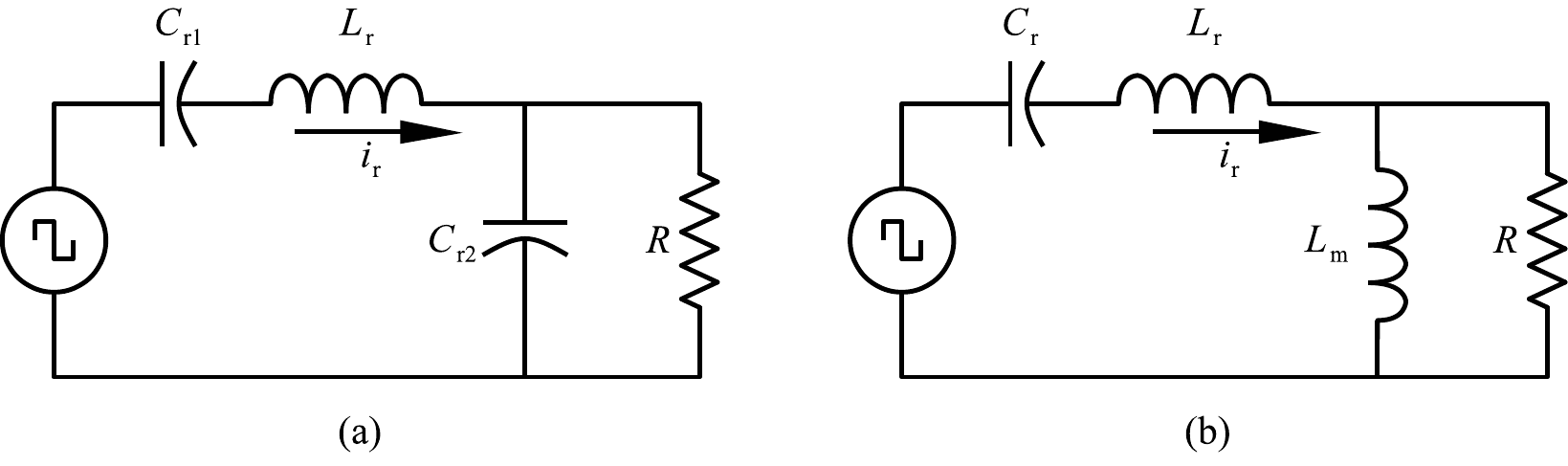}
	\caption{(a) LCC resonant converter; (b)  LLC resonant converter}
	\label{fig:Resonant02}
\end{figure}

To overcome the limitations of the resonant converters shown in \Fref{fig:Resonant01}, converters combining series and parallel topologies have been proposed. Figure \ref{fig:Resonant02}(a) shows a topology which includes two capacitors and one inductor. This topology is named the LCC resonant converter.

The topology formed by two inductors and one capacitor (the LLC resonant converter) is shown in \Fref{fig:Resonant02}(b). This topology has several advantages over the LCC topology. For instance, the two inductors can be integrated into one physical magnetic component, and it also provides galvanic isolation.

\begin{figure}
	\centering
	\includegraphics[width=0.9\linewidth]{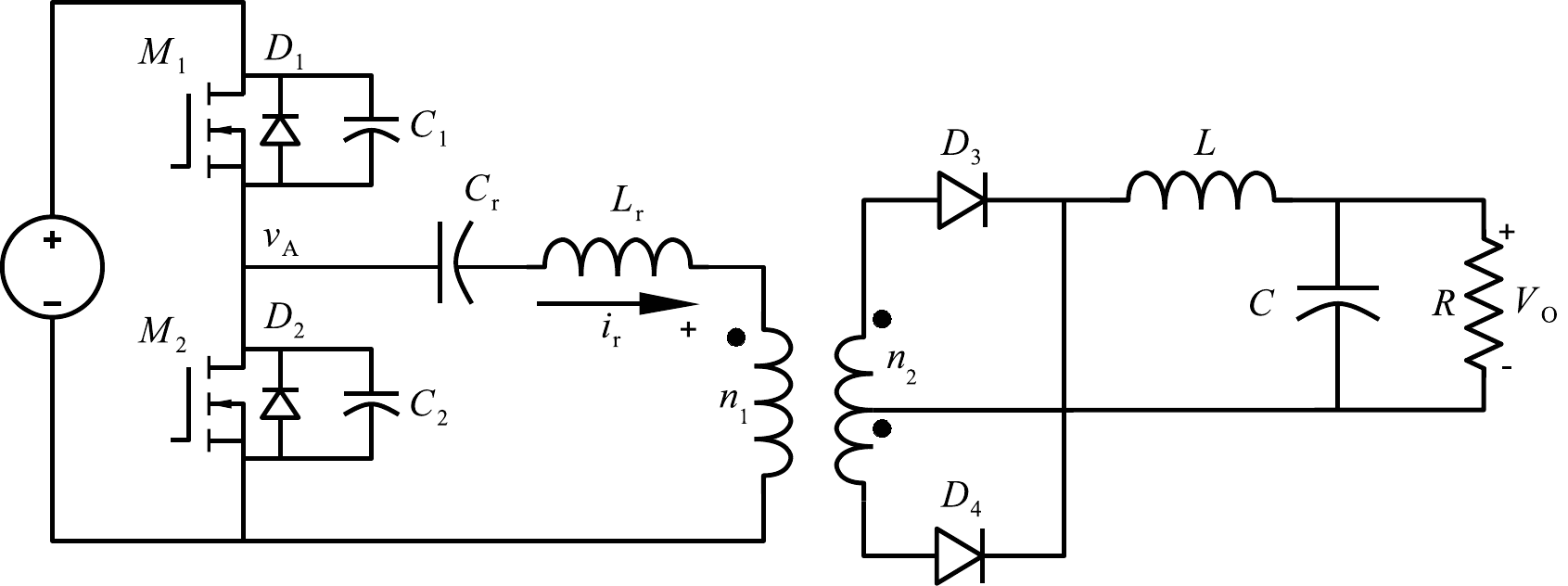}
	\caption{LLC resonant converter: circuit diagram}
	\label{fig:Resonant03}
\end{figure}

A simplified circuit diagram of an LLC resonant converter is shown in \Fref{fig:Resonant03}. The two inductors are the magnetizing and leakage inductances of a transformer. The LLC topology has the additional advantage of being able to operate in a soft switching mode. This allows the switching frequency and power density of the converter to be increased.

\begin{figure}
\centering
\includegraphics[scale=0.8]{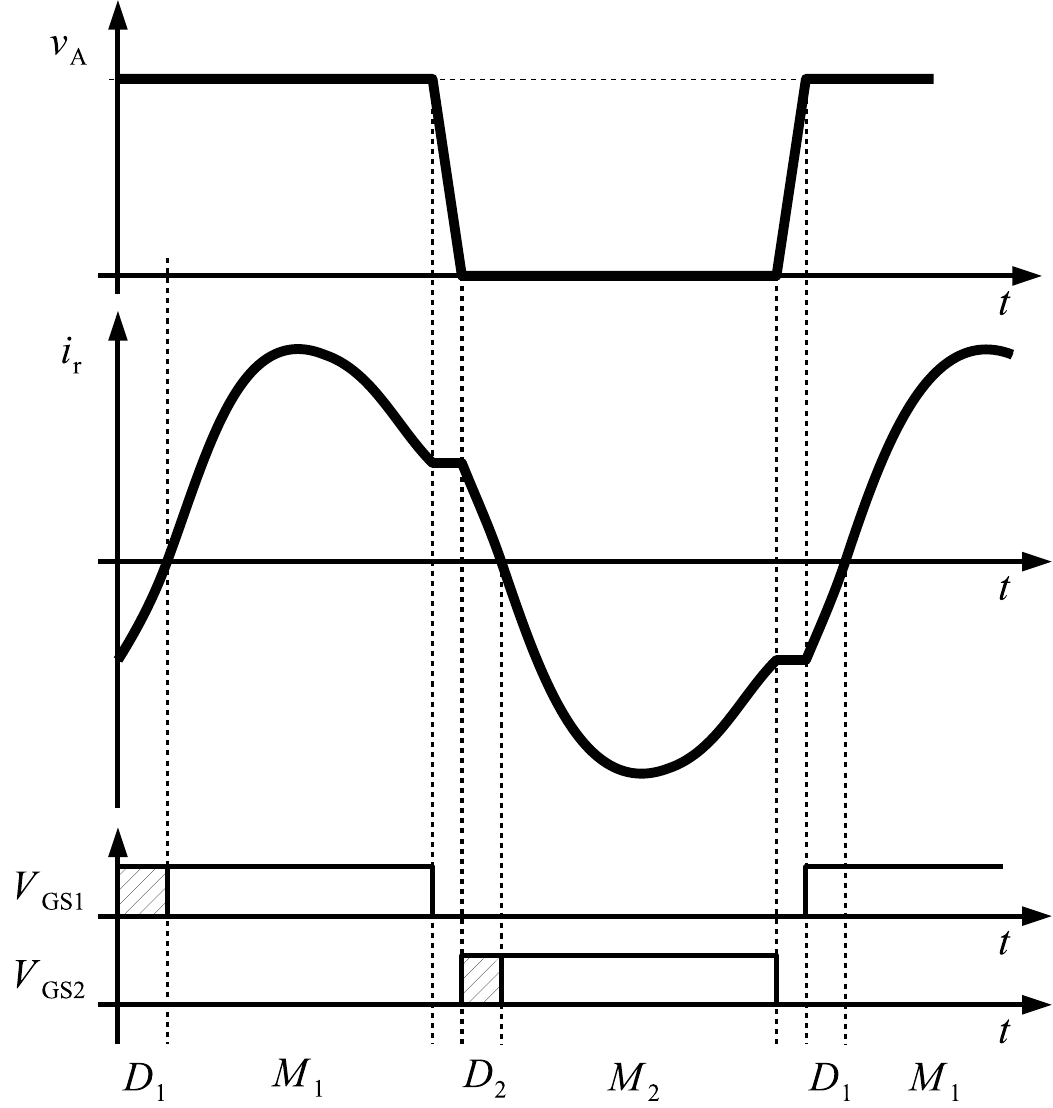}
\caption{Waveforms for LLC resonant converter}
\label{fig:Resonant04}
\end{figure}

The typical waveforms for an LLC resonant converter are shown in \Fref{fig:Resonant04}. Soft commutation is achieved in a similar way to that described in \Fref{fig:SoftSwitching02}.

\section{Conclusions}

One-quadrant power converters play an important role in industrial and consumer electronics applications. Almost every electronic circuit is supplied with a DC voltage provided by some kind of power converter. If a regulated voltage is needed, this converter usually consists of a rectifier stage followed by a regulated DC--DC power converter. The field of power converters for particle accelerators is no exception to this trend. As was mentioned in the introduction, one-quadrant power converters are used to generate the high-precision, stable currents needed for storage ring accelerator magnets. The topologies used in this application depend on the power level required.

This article was intended to be an introduction to this broad subject. All the topics discussed here are covered in depth in books on power converters and in journal and conference papers. The references provided at the end of this article should be a good starting point for further studies.

\nocite{rashid2001power}
\nocite{mohan2007power}
\nocite{bordry2004power}
\nocite{sheehan2007understanding}

\nocite{hung1993variable}
\nocite{tan2008general}

\nocite{cuk1983new}

\nocite{huang2010designing}

\nocite{Mappus2010}
\nocite{hegarty2007benefits}


\end{document}